\documentclass[prd,notitlepage,longbibliography,nofootinbib,superscriptaddress,twocolumn,preprintnumbers]{revtex4-2}
\usepackage[utf8]{inputenc}

\usepackage{bm}
\usepackage{comment} 
\usepackage[colorlinks=true,urlcolor=blue,anchorcolor=black,citecolor=blue,linkcolor=black,filecolor=black,menucolor=black,pagecolor=black,linktocpage=true,pdfproducer=medialab,pdfa=true]{hyperref}
\usepackage{graphicx}
\usepackage{amsmath,latexsym,amssymb,mathrsfs,ascmac,physics,color,subcaption}
\usepackage{multirow}
\usepackage{braket}
\usepackage[justification=raggedright]{caption}

\definecolor{DarkBlue}{rgb}{0,0,0.9} 

\definecolor{DarkRed}{rgb}{0.65,0,0}

\begin{document}
\title{
Removing naked singularities in static axially symmetric spacetimes by patching with the flat spacetimes}

\author{Daiki Saito}
\email{saito.daiki.g3@s.mail.nagoya-u.ac.jp}
\affiliation{Division of Science, Graduate School of Science, Nagoya University, Nagoya 464-8602, Japan}

\author{Daisuke Yoshida}
\email{dyoshida@math.nagoya-u.ac.jp}
\affiliation{Department of Mathematics, Nagoya University, Nagoya 464-8602, Japan}

\begin{abstract}
We investigate static, axially symmetric spacetimes without naked singularities that are constructed by patching Weyl class spacetimes with the flat spacetimes.
Once the exterior geometry is specified, the junction conditions determine the shape of a thin shell, which is the boundary between the two patched spacetimes, and the distribution of the energy and pressure on this shell, leaving a parameter representing the shell size free. 
We examine the cases where the exterior of the shell is given by Curzon--Chazy or Zipoy--Voorhees spacetimes. 
For each case, we find a lower bound on the shell size.
Additionally, we find that the weak and null energy conditions are satisfied for any shell size, while the dominant energy condition is satisfied for sufficiently large shells.
These results provide concrete examples of non-singular spacetimes with non-spherically symmetric exteriors that respect the energy conditions. 
\end{abstract}
\maketitle

\section{Introduction}
The black hole uniqueness theorem~\cite{Israel:1967wq} states that any asymptotically flat, static black hole solution to the vacuum Einstein equations must be a Schwarzschild spacetime with a positive Arnowitt--Deser--Misner (ADM) mass~\cite{Arnowitt:1960es}.
In other words, singularities in asymptotically flat, static, non-spherically symmetric solutions, as well as those in Schwarzschild spacetimes with a negative ADM mass, must be naked.
Since singularities indicate a breakdown of general relativity, the spacetime regions near the singularities are not reliable.
However, this does not necessarily imply that the regions far from the singularities are unrealistic, as these regions could potentially describe exteriors of regular spacetimes, such as stars.
The purpose of this paper is to explore the boundary between realistic and unrealistic regions near naked singularities from the perspective of the energy conditions for stars compensating for the singularities.
We focus on the null energy condition, the weak energy condition, and the dominant energy condition, which is expected to be satisfied by viable matter at least at the classical level.

Based on the positive mass theorem~\cite{Schoen:1979rg, Schoen:1981vd, Witten:1981mf}, which guarantees the non-negativity of the ADM mass under the dominant energy condition, we can immediately conclude that a Schwarzschild spacetime with a negative ADM mass cannot represent the exterior of a physically viable star due to the violation of the dominant energy condition. Consequently, any region in a Schwarzschild spacetime with a negative ADM mass is unrealistic, even if it is far from the naked singularity.

A class of exact, asymptotically flat solutions to the vacuum Einstein equations can be constructed if the spacetime possesses both static and axial symmetry. This class is known as the Weyl class \cite{Weyl:1917gp}. 
For the details of the Weyl class, see, for example, the textbooks~\cite{Stephani:2003tm, Griffiths:2009dfa}, as well as the references therein.
Though spacetimes in the Weyl class contain naked singularities in general, there is a sharp contrast to a Schwarzschild spacetime with a negative ADM mass: The ADM mass could be positive. Hence, the violation of the energy condition is not required by the positive mass theorem. Consequently, we expect that the naked singularities in spacetimes of the Weyl class exhibit different properties compared to those in Schwarzschild spacetimes with a negative ADM mass. In this paper, we focus on spacetimes in the Weyl class with a positive ADM mass only.

There are no general guidelines for assessing the viability of energy conditions for stars that compensate for the naked singularities in the Weyl class, such as the positive mass theorem for Schwarzschild spacetime with a negative ADM mass.
In this paper, we investigate this issue by using a simple model of a star: an axisymmetric thin shell surrounding flat spacetime.
As is well known, the dynamics of a thin shell are described by the junction conditions between the interior and exterior spacetimes~\cite{Israel:1966rt}. 
Unlike many studies on thin shells that focus on the motion of a shell with a given equation of state, such as Refs.~\cite{Lake:1979zz,Lake:1980zz,Siegel:1981xd,Kodama:1981gu,Berezin:1982ur,Ipser:1983db,Berezin:1983dz,Hoye:1983dq,Maeda:1985ye,Sato:1986,Berezin:1987bc} for spherically symmetric case and Refs.~\cite{DeLaCruz:1968zz,Lindblom:1974bq,Pereira:2000tc,Goncalves:2002my,Mann:2008rx,Delsate:2014iia,Nakao:2014qva, Ogasawara:2018gni,Oshita:2019jan,Saito:2021vut} for axisymmetric rotating case, we do not assume an equation of state \textit{a priori}.
Instead, we require that the shell's motion be static.
Then the junction conditions provide the conditions for the shell's position, energy density, and pressure necessary to achieve the required staticity.
Although the actual physical properties of this shell are not fully clear, we can at least examine the energy conditions.
This approach allows us to explore the boundary between realistic and unrealistic regions based on criteria related to the energy conditions. For the spherically symmetric case, a similar approach is studied in Ref.~\cite{Frauendiener:1990nao}, where it is found that the dominant energy condition is violated if the areal radius of the shell is smaller than $\frac{25}{24}$ times the Schwarzschild radius.
The method of removing singularities by patching spacetimes with localized energy has also been studied in Ref.~\cite{Conboy:2005nx}, although it focuses on temporally localized energy rather than spatially localized energy, such as shells.

This paper is organized as follows. In the next section, we review the basics of static, axisymmetric solutions to the Einstein equations. In Sec.~\ref{junc}, we derive the junction conditions between two Weyl-class solutions separated by a static, axisymmetric shell.
Then, in Sec.~\ref{exp}, we investigate two classes of exact solutions within the Weyl class as the exterior of the shell: 
the Curzon--Chazy spacetimes~\cite{Curzon:1925, Chazy:1924} and the Zipoy--Voorhees spacetimes \cite{Zipoy:1966btu, Voorhees:1970ywo}.
In Sec.~\ref{exp}, we derive the shape, energy, and pressure of the shell from the junction conditions and examine the minimum size of the shell in light of the energy conditions for each case.
The Sec.~\ref{summary} provides a summary and discussion. Throughout this paper, we use the boldstyle notation basically following Misner--Thorne--Wheeler's textbook \cite{Misner:1973prb}.
We also use a shorthand notation $\boldsymbol{u}\boldsymbol{v} := \frac{1}{2} \left( \boldsymbol{u} \otimes \boldsymbol{v} + \boldsymbol{v} \otimes \boldsymbol{u} \right)$ for any tensors $\boldsymbol{u}$ and $\boldsymbol{v}$, and, especially, if $\boldsymbol{u} = \boldsymbol{v}$, we simply express this as $\boldsymbol{u}^2 := \boldsymbol{u} \boldsymbol{u} = \boldsymbol{u} \otimes \boldsymbol{u}$.
Throughout this paper, we use a unit in which the speed of light is unity.

\section{Static Axially Symmetric Spacetimes}
\label{SASS}

In this section, we will review the basic properties of static, axially symmetric spacetimes. For a more detailed discussion, see, for example,  the textbooks~\cite{Stephani:2003tm, Griffiths:2009dfa}.

Let $({\cal M}, \boldsymbol{g})$ be a static, axially symmetric spacetime. By static and axial symmetries, the spacetime possess the two Killing vectors, $\boldsymbol{t} = t^{\mu} \boldsymbol{\partial}_{\mu}$ and $\boldsymbol{\phi} = \phi^{\mu}\boldsymbol{\partial}_{\mu}$, which are hypersurface orthogonal and commute with each other. That is to say, $\boldsymbol{t}$ and $\boldsymbol{\phi}$ satisfy the Killing equations:
\begin{align}
\mathsterling_{\boldsymbol{t}} \boldsymbol{g} = 0, \qquad \mathsterling_{\boldsymbol{\phi}} \boldsymbol{g} = 0,\label{eq:Killingeqs} 
\end{align}
and the hypersurface orthogonal conditions:
\begin{align}
t_{[\mu} \nabla_{\nu} t_{\rho]} = 0, \qquad \phi_{[\mu} \nabla_{\nu} \phi_{\rho]} = 0,\label{eq:twist-free}
\end{align}
and the commutation relation:
\begin{align}
 [\boldsymbol{t}, \boldsymbol{\phi}] = 0. \label{eq:commutation}
\end{align}
By the commutation relation~\eqref{eq:commutation}, we can introduce the coordinates so that $\boldsymbol{t}$ and $\boldsymbol{\phi}$ form two of the coordinate basis: $\boldsymbol{t} = \boldsymbol{\partial}_{t}$ and $\boldsymbol{\phi} = \boldsymbol{\partial}_{\phi}$. In addition, by the hypersurface orthogonal conditions~\eqref{eq:twist-free}, we can assign the remaining two coordinates, say, $\eta$ and $\xi$, as coordinates on the hypersurface orthogonal to the two Killing vectors. In addition, since this hypersurface is two-dimensional, we can choose $\eta$ and $\xi$ as conformal flat coordinates without loss of generality. Hence the metric can be expressed as
\begin{align}
\boldsymbol{g} = - \mathrm{e}^{2 U} \boldsymbol{d}t^2 + \mathrm{e}^{-2U} \qty(\rho^2 \boldsymbol{d}\phi^2 + \mathrm{e}^{2\gamma}\left(\boldsymbol{d}\eta^2 + \boldsymbol{d}\xi^2 \right)). \label{eq:dsWeyl}
\end{align}
Because of the symmetry \eqref{eq:Killingeqs}, the functions $U$, $\gamma$ and $\rho$ depend only on $\eta$ and $\xi$.

Using this metric ansatz, one of the vacuum Einstein equations $R_{\mu\nu}=0$ reads
\begin{align}
   &\partial^2_{\eta}\rho+\partial^2_{\xi}\rho = 0, \label{eq:rho}
\end{align}
which indicates that $\rho$ is a harmonic function.
In this paper, we adopt $\rho$ and its harmonic conjugate $z$, which is defined by $ \partial_{\eta} z = - \partial_{\xi} \rho$ and $\partial_{\xi} z = \partial_{\eta} \rho$, as the coordinates, which are known as Weyl's canonical coordinates.
In terms of these coordinates, the metric can be written as
\begin{align}
\boldsymbol{g} = - \mathrm{e}^{2 U} \boldsymbol{d}t^2 + \mathrm{e}^{-2U} \qty[\rho^2 \boldsymbol{d}\phi^2 + \mathrm{e}^{2\gamma}\left(\boldsymbol{d}\rho^2 + \boldsymbol{d}z^2 \right)], \label{eq:dsW}
\end{align}
where $U$ and $\gamma$ now depend on $\rho$ and $z$.
Then, the remaining components of the Einstein equation can be rewritten as
\begin{align}
 &\partial^2_{\rho}U+\partial^2_{z}U+\frac{\partial_{\rho}U}{\rho} = 0, \label{eq:Lap} \\
 &\partial_{\rho}\gamma = \rho\qty(\qty(\partial_{\rho}U)^2-\qty(\partial_{z}U)^2), \label{eq:gammarho} \\
 &\partial_{z}\gamma = 2\rho\partial_{\rho}U\partial_{z}U. \label{eq:gammaz}
\end{align}
Once we solve Eq.~\eqref{eq:Lap} to obtain $U$, the function $\gamma$ can be obtained by integrating Eqs.~\eqref{eq:gammarho} and \eqref{eq:gammaz}, which are integrable.
The key observation is that Eq.~\eqref{eq:Lap} is nothing but the Laplace equation for an axisymmetric potential in the three-dimensional Euclidean space,
which can be confirmed explicitly by introducing asymptotic Cartesian coordinates $x^{i} = \{x,y,z\}$ through
\begin{align}
    x = \rho \cos \phi, \qquad y = \rho \sin \phi,
\end{align}
and rewriting Eq.~\eqref{eq:Lap} as
\begin{align}
 \left( \partial_{x}^2 + \partial_{y}^2 + \partial_{z}^2 \right) U\qty(\sqrt{x^2 + y^2}, z) = 0.
\end{align}
Hence, we can solve it for a given boundary condition as is the Laplace equation for the Newton potential.

The trivial solution $U=0$ and $\gamma=0$ corresponds to the Minkowski spacetime, though other choices for describing Minkowski spacetime, such as $U = \gamma = \log \rho$, are also possible. The general class of the asymptotically flat solutions of Eq.~\eqref{eq:Lap} can be obtained in terms of the multipole expansions,
\begin{align}
U = -\sum_{i=0}^{\infty} \frac{a_{i}}{|\vec{x}|^{i+1}} P_{i}\left( \frac{z}{|\vec{x}|}\right), \label{eq:WeylU}
\end{align}
which is known as the Weyl class. Here, 
$|\vec{x}| := \sqrt{x^2 + y^2 +z^2} = \sqrt{\rho^2 + z^2}$, $P_{i}$ are Legendre polynimials, and $a_{i}$ are arbitrary constants.
By integrating Eqs.~\eqref{eq:gammarho} and \eqref{eq:gammaz}, $\gamma$ can be obtained as
\begin{align}
 &\gamma =-\sum_{j,k=0}^{\infty}a_{j}a_{k}\frac{(j+1)(k+1)}{j+k+2}\frac{P_{j}P_{k}-P_{j+1}P_{k+1}}{|\vec{x}|^{j+k+2}}. \label{eq:Weylgamma}
\end{align}
The leading order of the asymptotic expansions is characterized by the constant $a_{0}$.
The metric can be expanded as
\begin{align}
 \boldsymbol{g} &\sim - \left( 1 - \frac{2 a_{0}}{|\vec{x}|} \right) \boldsymbol{d}t^2 + \left( 1 + \frac{2 a_{0}}{|\vec{x}|} \right) \delta_{ij} \boldsymbol{d}x^{i} \boldsymbol{d}x^{j},
\end{align} 
where we have omitted the $\mathcal{O}\left( 1/|\vec{x}|^2 \right)$ contributions.
From this expression, the constant $a_0$ can be understood as the ADM mass multiplied by the Newton constant $G$. In other words, the monopole component of $U$ represents the monopole component of the actual Newton potential in the asymptotic region.
See also Refs.~\cite{1990ForPh..38..733Q,1994GReGr..26..877H} for the discussion on multipole moments of a source in an asymptotically flat spacetimes.

In this paper, we mainly focus on two simple classes in the Weyl class: Curzon--Chazy spacetimes~\cite{Curzon:1925, Chazy:1924} and Zipoy--Voorhees spacetimes \cite{Zipoy:1966btu, Voorhees:1970ywo} as described below:

\subsection{Curzon--Chazy spacetime}
A simple, non-trivial solution of the Laplace equation \eqref{eq:Lap} is 
obtained with the boundary condition which corresponds to introducing the point-like source at the origin $\rho = z = 0$. The asymptotically flat solution is the Newton potential
\begin{align}
 U(\rho, z) = - \frac{G m}{\sqrt{\rho^2 + z^2}},\label{eq:CCU}
\end{align}
where we introduce a free parameter $m$ with the dimension of mass. By integrating Eqs.~\eqref{eq:gammarho} and \eqref{eq:gammaz}, $\gamma$ can be obtained as
\begin{align}
 \gamma(\rho, z) = - \frac{G^2 m^2 \rho^2}{2 (\rho^2 + z^2)^2}.\label{eq:CCgamma}
\end{align}
Here we fix the integration constant so that $\gamma \rightarrow 0$ at asymptotic infinity. We note that, although the function $U$ is spherically symmetric, $\gamma$ is not. Hence the spacetime is not spherically symmetric.

To summarize, the static, axisymmetric spacetime~\eqref{eq:dsWeyl} with the functions $U$ and $\gamma$ specified by Eqs.~\eqref{eq:CCU} and \eqref{eq:CCgamma} is a solution to the vacuum Einstein equations with one free parameter $m$, which is known as Curzon--Chazy spacetime \cite{Curzon:1925, Chazy:1924}. This solution corresponds to $a_{0} = G m$, $a_{i \geq 1} = 0$ in the general Weyl class. Hence the parameter $m$ describes the ADM mass of this spacetime and we focus on the case with $m > 0$.

The curvature can be evaluated as
\begin{widetext}
\begin{align}
 C_{\mu\nu\rho\sigma} C^{\mu\nu\rho\sigma}
= \mathrm{e}^{- \frac{4 G m}{|\vec{x}|}\left(1 - \frac{G m \rho^2}{2 |\vec{x}|^{3}} \right)}
\frac{48 G^2 m^2}{|\vec{x}|^6}
\qty(
1
- \frac{2G m}{|\vec{x}|}
+ \frac{G^2 m^2 }{|\vec{x}|^2}
+ \frac{G^2 m^2 \rho^2}{|\vec{x}|^4}
- \frac{G^3 m^3 \rho^2}{|\vec{x}|^5}
+ \frac{G^4 m^4 \rho^2}{3 |\vec{x}|^6}
),
\end{align} 
\end{widetext}
with $C_{\mu\nu\rho\sigma}$ denotes the components of the Weyl tensor.
Since it diverges in the limit $\rho \rightarrow 0, z \rightarrow 0$, the origin $\rho = z = 0$ is the scalar curvature singularity. 
As there is no event horizon, this singularity is naked. 
For a more detailed discussion on this singularity, such as the directional dependence, see Refs.~\cite{STACHEL:196860,Gautreau:1967,Cooperstock:1974,Morgan:1973va,scott1986curzon1,scott1986curzon2,Abdelqader:2012ey}.

\subsection{Zipoy--Voorhees spacetime}
Let us consider yet another class within the Weyl class where $U$ corresponds to the Newton potential sourced by a rod with the length $2 \ell > 0$ and the mass linear density $m/ 2\ell$, which is located at $\rho = 0$ and $z \in [ -\ell, \ell ]$.
Such Newton potential can be evaluated as 
\begin{align}
 U(\rho,z) &= - \int_{- \ell}^{\ell} \frac{G m}{\sqrt{\rho^2 + (z - \zeta)^2}} \frac{d \zeta}{2 \ell} \notag \\
& = \frac{G m}{2 \ell} \log \left(
\frac{\sqrt{\rho^2 + (z - \ell)^2} + z - \ell}{\sqrt{\rho^2 + (z + \ell)^2} + z + \ell}
\right). \label{eq:UZV}
\end{align}
Again, by integrating Eqs.~\eqref{eq:gammarho} and \eqref{eq:gammaz}, $\gamma$ can be obtained as
\begin{align}
& \gamma(\rho, z) = \frac{G^2 m^2}{2 \ell^2} \log \Xi(\rho,z), \label{eq:gammaZV}
\end{align}
where $\Xi$ is defined by
\begin{align}
\Xi :=  \frac{\left( \sqrt{\rho^2 + (z + \ell)^2} + \sqrt{\rho^2 + (z - \ell)^2} \right)^2 - 4 \ell^2}{4 \sqrt{\rho^2 + (z + \ell)^2} \sqrt{\rho^2 + (z - \ell)^2}}.
\end{align}
These two parameter families, specified by $m$ and $\ell$, of the solutions to the vacuum Einstein equations, are known as Zipoy--Voorhees spacetime~\cite{Zipoy:1966btu, Voorhees:1970ywo}.
The Zipoy--Voorhees class corresponds to the Weyl class with the parameter choice
\begin{align}
 a_{i} = 
\begin{cases}
0 & (i \text{: odd}) \\
\frac{G m}{i + 1} \ell^{i}  & (i \text{: even})
\end{cases}.
\end{align}
In particular, $m$ corresponds to the ADM mass because $a_{0} = G m$.

There are two important choices of the parameters.
One is $\ell = G m$, where the metric is known to describe the Schwarzschild spacetime outside the black hole/white hole horizon. The other is $\ell \rightarrow 0$, where the metric describes the Curzon--Chazy class.

The metric components in Weyl's canonical coordinates are singular at $\rho = 0$ with $z \in [ -\ell, \ell ]$.
By expanding the curvature around $\rho = 0$ with fixing $z \in (-\ell, \ell)$, 
we obtain
\begin{widetext}
\begin{align}
&C_{\mu\nu\rho\sigma} C^{\mu\nu\rho\sigma} \notag\\
& = 
\frac{1}{\ell^4} \left( \frac{G m}{\ell}\right)^2 \left(1 - \frac{G m}{\ell}\right)^2
\left(1 - \frac{G m}{\ell} + \left(\frac{G m}{\ell}\right)^2 \right) \frac{\rho^{- 4 \left(1 - \frac{G m}{\ell}  + \left(\frac{G m}{\ell}\right)^2 \right)}}{(2 \ell)^{- 4}
\left( 2 \sqrt{\ell^2 - z^2}\right)^{4 \frac{G m}{\ell}}
 (\frac{\ell^2 - z^2}{\ell})^{- 4 \left(\frac{G m}{\ell}\right)^2} } \left( 1 + {\cal O}\qty(\rho^2)\right).
\end{align} 
\end{widetext}
Since $ 1 - \frac{G m}{\ell} + \left(\frac{G m}{\ell}\right)^2 \geq \frac{3}{4}$,
the Weyl square diverges in the limit $\rho \rightarrow 0$ at least as fast as $\rho^{-3}$ except when $\ell = G m$. Thus, for $\ell \neq G m$, the location $\rho = 0$ and $z \in [ -\ell, \ell ]$ is a scalar curvature singularity. For $\ell = G m$, the curvature vanishes, and the location $\rho = 0$ with $z \in [ -\ell, \ell ]$ corresponds to a coordinate singularity at the horizon.
For a more detailed property of this solution, see the related works such as Refs.~\cite{Gautreau:1967,Bonnor:1968,Esposito:1975qkv,Kodama:2003ch,Hoenselaers:1978,Papadopoulos:1981wr,Stewart:1982}.

\section{Junction Conditions}
\label{junc}

\subsection{The first junction conditions}
Let us consider a spacetime $(\cal M, \boldsymbol{g})$ where two static, axially symmetric spacetimes, ${\cal M}^{+}$ and ${\cal M}^{-}$, are glued through an axially symmetric shell with a world volume $\Sigma$. In the following part of the paper, the indices $+/-$ denote the quantities in the outside/inside of the shell.

Contrary to the spherically symmetric case where Birkhoff's theorem holds, we need to assume that the shell is static. Hence, our spacetime is globally static. Let us write the global Killing vectors associated with the time translation and the axial rotation by $\boldsymbol{t}$ and $\boldsymbol{\phi}$, respectively. In each spacetime region ${\cal M}^{\pm}$, we introduce Weyl's canonical coordinates $\{t_{\pm}, \phi_{\pm}, \rho_{\pm}, z_{\pm} \}$ so as the global Killing vector fields and the metric are expressed as 
\begin{align}
 \boldsymbol{t} &= 
\begin{cases}
 c_{+} \boldsymbol{\partial}_{t_{+}} & \text{in }{\cal M}^{+} \\
 c_{-} \boldsymbol{\partial}_{t_{-}} & \text{in }{\cal M}^{-}
\end{cases},
 \\
\boldsymbol{\phi} &=
\begin{cases}
  \boldsymbol{\partial}_{\phi_{+}} & \text{in }{\cal M}^{+} \\
  \boldsymbol{\partial}_{\phi_{-}} & \text{in }{\cal M}^{-}
\end{cases},
\end{align}
and
\begin{widetext}
\begin{align}
g_{\mu\nu} \boldsymbol{d}x^{\mu} \boldsymbol{d}x^{\nu} = 
\begin{cases}
- \mathrm{e}^{2 U_{+}} \boldsymbol{d}t_{+}^2 + \mathrm{e}^{ - 2 U_{+}} \qty(\rho_{+}^2 \boldsymbol{d} \phi_{+}^2 + \mathrm{e}^{2 \gamma_{+}} \left(\boldsymbol{d} \rho_{+}^2 + \boldsymbol{d} z_{+}^2 \right)) & \text{in }{\cal M}^{+} \\
- \mathrm{e}^{2 U_{-}} \boldsymbol{d}t_{-}^2 + \mathrm{e}^{ - 2 U_{-}} \qty(\rho_{-}^2 \boldsymbol{d} \phi_{-}^2 + \mathrm{e}^{2 \gamma_{-}} \left(\boldsymbol{d} \rho_{-}^2 + \boldsymbol{d} z_{-}^2 \right))
& \text{in }{\cal M}^{-}
\end{cases},
\end{align}
\end{widetext}
respectively.
Here, $c_{\pm}$ are positive constants and $U_{\pm}$ and $\gamma_{\pm}$ are given functions of $\rho_{\pm}$ and $z_{\pm}$. 

Let us describe the world volume of the shell as the three-dimensional sub-manifold $\left( \Sigma, \boldsymbol{h} \right)$ embedded into ${\cal M}$.
Note that we express both the three-dimensional manifold and its image embedded into ${\cal M}$ by $\Sigma$. 
Let us introduce the coordinates of $\Sigma$ by $\qty{y^{i}} = \{\tau, \varphi, r\}$. We denote the embedding of $\Sigma$ into the spacetime ${\cal M}$ by the map
\begin{align}
 \{\tau, \varphi, r\}
&\mapsto
\{t_{\pm}, \phi_{\pm}, \rho_{\pm}, z_{\pm} \} \notag\\
& \qquad = \qty{ c_{\pm} \tau, \varphi ,\frac{1}{c_{\pm}} R_{\pm}(r) , \frac{1}{c_{\pm}} Z_{\pm}(r)}.
\end{align}
We express the pullback associated with this embedding by $|_{\Sigma}$. 
Thus, the induced metric, the pullback of $\boldsymbol{g}$ from each spacetime region ${\cal M}_{\pm}$ can be evaluated as 
\begin{align}
\boldsymbol{g}|_{\Sigma}
 &= - c_{\pm}^2 \mathrm{e}^{2 U_{\pm}} \boldsymbol{d}\tau^2 \nonumber \\
 &+ \frac{1}{c_{\pm}^2 \mathrm{e}^{2 U_{\pm}}} \qty( R_{\pm}^2 \boldsymbol{d}\varphi^2 + \mathrm{e}^{2 \gamma_{\pm}} (R_{\pm}'(r)^2 + Z_{\pm}'(r)^2 ) \boldsymbol{d}r^2 ). \label{eq:inmet}
\end{align}
The first junction condition can be derived by requiring that the expressions for the induced metric Eq.~\eqref{eq:inmet} coincide for the outside and the inside of the shell:
\begin{align}
c_{+}^2 \mathrm{e}^{2 U_{+}(r)} &= c_{-}^2 \mathrm{e}^{2 U_{-}(r)}, \label{eq:1stf} \\
R_{+}(r) &= R_{-}(r), \label{eq:1stR}
\end{align}
and
\begin{align}
&\mathrm{e}^{\gamma_{+}(r)} \sqrt{R'_{+}(r)^2 + Z'_{+}(r)^2}  \notag\\
&\qquad = \mathrm{e}^{\gamma_{-}(r)} \sqrt{R'_{-}(r)^2 + Z'_{-}(r)^2}. \label{eq:1sta}
\end{align}
Here $U_{\pm}(r), \gamma_{\pm}(r)$ in above expressions are understood as
\begin{align}
U_{\pm}(r) &=  U_{\pm}\qty(\frac{R_{\pm}(r)}{c_{\pm}}, \frac{Z_{\pm}(r)}{c_{\pm}}), \\
\gamma_{\pm}(r) &=  \gamma_{\pm}\qty(\frac{R_{\pm}(r)}{c_{\pm}}, \frac{Z_{\pm}(r)}{c_{\pm}}).
\end{align}
By denoting the right-hand sides of  Eqs. \eqref{eq:1stf}, \eqref{eq:1stR}, and \eqref{eq:1sta} as $f(r), R(r)$ and $a(r)$ respectively, the induced metric $\boldsymbol{g}|_{\Sigma}$ can be expressed as 
\begin{align}
\boldsymbol{g}|_{\Sigma} = - f(r) \boldsymbol{d} \tau^2 + \frac{1}{f(r)} \left(  R(r)^2 \boldsymbol{d} \varphi^2 + a(r)^2 \boldsymbol{d} r^2 \right).
\end{align}
We note that there is a freedom in choosing the $r$ coordinate on $\Sigma$. For instance, we can set $R(r) = r$ without loss of generality. 
\subsection{The second junction conditions}
The push-forward of the coordinate basis of $\Sigma$ can be evaluated as 
\begin{align}
\boldsymbol{\partial}_{\tau} &\mapsto \boldsymbol{\theta}_{\tau}:= c_{\pm} \boldsymbol{\partial}_{t_{\pm}} = \boldsymbol{t}, \\
\boldsymbol{\partial}_{\varphi} &\mapsto \boldsymbol{\theta}_{\varphi}:=\boldsymbol{\partial}_{\phi_{\pm}} = \boldsymbol{\phi}, \\
 \boldsymbol{\partial}_{r} &\mapsto \boldsymbol{\theta}_{r}:= \frac{1}{c_{\pm}} \left(R'_{\pm}(r) \boldsymbol{\partial}_{\rho_{\pm}} + Z'_{\pm}(r) \boldsymbol{\partial}_{z_{\pm}} \right)
\end{align}
which forms the basis of the vectors tangent to $\Sigma$ in ${\cal M}$.
Then one can define the unit normal to $\Sigma$ as the 1-form which is orthogonal to the above three vector fields:
\begin{align}
\boldsymbol{n}_{\pm} &:=  \epsilon_{\pm} \mathrm{e}^{- U_{\pm}(r) + \gamma_{\pm}(r)} \frac{ Z_{\pm}'(r) \boldsymbol{d}\rho_{\pm} - R_{\pm}'(r) \boldsymbol{d}z_{\pm} }{\sqrt{R_{\pm}'(r)^2 + Z_{\pm}'(r)^2}},
\end{align}
where $\epsilon_{\pm} = 1$ or $-1$.
By defining the vector
\begin{align}
 \boldsymbol{\theta}_{n} & := n_{\pm}^{\mu} \boldsymbol{\partial}_{\mu} \nonumber \\
 &= 
\epsilon_{\pm} \mathrm{e}^{U_{\pm}(r) - \gamma_{\pm}(r)} \frac{Z_{\pm}'(r) \boldsymbol{\partial}_{\rho_{\pm}} - R_{\pm}'(r) \boldsymbol{\partial}_{z_{\pm}}}{\sqrt{R_{\pm}'(r)^2 + Z_{\pm}'(r)^2}}
,
\end{align}
we may use $\{\boldsymbol{\theta}_{\mu}\} = \{ \boldsymbol{\theta}_{\tau}, \boldsymbol{\theta}_{\varphi}, \boldsymbol{\theta}_{r}, \boldsymbol{\theta}_{n} \}$ as the basis of the vector space.
We can also introduce the dual basis $\{ \boldsymbol{e}^{\mu} \} = \{ \boldsymbol{e}^{\tau}, \boldsymbol{e}^{\varphi}, \boldsymbol{e}^{r}, \boldsymbol{e}^{n} \}$ as
\begin{align}
 \boldsymbol{e}^{\tau} &= \frac{1}{c_{\pm}} \boldsymbol{d} t_{\pm}, \\
 \boldsymbol{e}^{\varphi} &= \boldsymbol{d} \phi_{\pm}, \\
 \boldsymbol{e}^{r} &= c_{\pm}\frac{R'(r) \boldsymbol{d}\rho_{\pm} + Z'(r) \boldsymbol{d}z_{\pm}}{R'(r)^2 + Z'(r)^2}, \\
 \boldsymbol{e}^{n} &= \boldsymbol{n},
\end{align}
so that they satisfy $\braket{\boldsymbol{\theta}^{\mu},\boldsymbol{e}_{\nu}} = \delta^{\mu}{}_{\nu}$.
Note that the pullback of these bases reduces to the coordinate basis on $\Sigma$:
\begin{align}
  \boldsymbol{e}^{\tau}|_{\Sigma} &= \boldsymbol{d} \tau, \\
 \boldsymbol{e}^{\varphi}|_{\Sigma} &= \boldsymbol{d}\varphi , \\
 \boldsymbol{e}^{r}|_{\Sigma} &= \boldsymbol{d} r, \\
 \boldsymbol{e}^{n}|_{\Sigma} &= 0.
\end{align}
We can also represent the induced metric as a degenerate tensor in $\mathcal{M}$ as 
\begin{align}
\boldsymbol{h} &= \boldsymbol{g} - \boldsymbol{n} \boldsymbol{n}\\
&= - f(r) \boldsymbol{e}^{\tau} \boldsymbol{e}^{\tau}
+ \frac{1}{f(r)} \left(R(r)^2 \boldsymbol{e}^{\varphi} \boldsymbol{e}^{\varphi} + a(r)^2 \boldsymbol{e}^{r} \boldsymbol{e}^{r} \right).
\end{align}
Clearly, the pullback of $\boldsymbol{h}$ reduces to 
the induced metric $\boldsymbol{g}|_{\Sigma}$ on $\Sigma$. 
Hence $\boldsymbol{h}$ is also referred to as the induced metric.

From the expression of $\boldsymbol{n}_{\pm}$, one can evaluate the extrinsic curvature $\boldsymbol{K}_{\pm} = K_{\pm\mu\nu} \boldsymbol{d}x^{\mu} \boldsymbol{d} x^{\nu}$ defined by $ K_{\pm\mu\nu} := h_{\mu}{}^{\rho} \nabla_{\rho} n_{\pm}{}_{\nu}$. 
We note that $\boldsymbol{n}_{\pm}$ is defined only on $\Sigma$ and hence, in order to evaluate the covariant derivative of $\boldsymbol{n}_{\pm}$, one needs to extend the definition of $\boldsymbol{n}_{\pm}$ to the neighborhood of $\Sigma$, though the expression of $\boldsymbol{K}_{\pm}$ is independent of the way to extend.
Then, the extrinsic curvature can be expressed as
\begin{align}
 \boldsymbol{K}_{\pm} = K_{\pm\tau\tau} \boldsymbol{e}^{\tau}\boldsymbol{e}^{\tau} + K_{\pm\varphi \varphi} \boldsymbol{e}^{\varphi} \boldsymbol{e}^{\varphi} + K_{\pm rr} \boldsymbol{e}^{r} \boldsymbol{e}^{r} .
\end{align}
Here the components and the trace, $ K_{\pm} := g^{\mu\nu} K_{\pm\mu\nu}$, are given by
\begin{align}
 K_{\pm\tau\tau} &= 
 \nabla_{n} U_{\pm} 
~ h_{\tau\tau}(r), \label{eq:Ktau}
\\
 K_{\pm\varphi\varphi} &= 
 \left(
- \nabla_{n} U_{\pm}
+ \epsilon_{\pm} \frac{\sqrt{f(r)}}{a(r)}  \frac{Z'_{\pm}}{R}
\right)
h_{\varphi\varphi}(r), \label{eq:Kphi}
\end{align}
and
\begin{widetext}
\begin{align}
 K_{\pm rr} &= 
\biggl(
 \nabla_{n} ( - U_{\pm} + \gamma_{\pm})
+ \epsilon_{\pm} \frac{\sqrt{f(r)}}{a(r)}  \frac{R'(r) Z_{\pm}''(r) - Z_{\pm}'(r) R''(r)}{R'(r)^2 + Z_{\pm}'(r)^2}
 \biggr) h_{rr}(r), \label{eq:Kr} \\
 K_{\pm} &= 
\nabla_{n}(- U_{\pm} + \gamma_{\pm}) +  \epsilon_{\pm} \frac{\sqrt{f(r)}}{a(r)}  \left(\frac{Z_{\pm}'(r)}{R(r)} +  \frac{R'(r) Z_{\pm}''(r) - Z_{\pm}'(r) R''(r)}{R'(r)^2 + Z_{\pm}'(r)^2}\right) , \label{eq:Ktr}
\end{align}    
\end{widetext}
where $\nabla_{n} := \nabla_{\boldsymbol{\theta}_{n}}$ and it can be evaluated as
\begin{align}
\nabla_{n} F = \epsilon_{\pm} \frac{\sqrt{f(r)}}{c_{\pm} a(r)} \left(
 Z_{\pm}'(r) \partial_{\rho} F - R'(r) \partial_{z} F
\right),
\end{align}
for any scalar function $F$.

The second junction condition for a vacuum solution of the Einstein equation can be summarized as 
\begin{align}
 & \left[K_{\mu\nu} - h_{\mu\nu} K \right]^{-} = - 8 \pi G S_{\mu\nu}, \label{eq:2nd1}\\
 & [K_{\mu\nu}]^{+} S^{\mu\nu} = 0
 \label{eq:2nd2}, \\
 & D_{\nu} S^{\nu}{}_{\mu} = 0
\label{eq:2nd3}
,
\end{align}
where we have introduced the notation,
\begin{align}
 [F]^{-} := F_{+} - F_{-}, \qquad  [F]^{+} := F_{+} + F_{-}.
\end{align}
The tensor $\boldsymbol{S} = S_{\mu\nu} \boldsymbol{d}x^{\mu} \boldsymbol{d}x^{\nu}$ is the energy momentum tensor localized on $\Sigma$.
In this paper, we will assume that the shell has energy and anisotropic tangential pressures:
\begin{align}
    \boldsymbol{S} &= - \varepsilon(r) h_{\tau\tau}(r) \boldsymbol{e}^{\tau} \boldsymbol{e}^{\tau}
    \notag \\
    & \qquad
+ p_{\varphi}(r) h_{\varphi\varphi}(r) \boldsymbol{e}^{\varphi} \boldsymbol{e}^{\varphi}
    +p_{r}(r) h_{rr}(r) \boldsymbol{e}^{r} \boldsymbol{e}^{r}. \label{eq:S}
\end{align}
Here, $\varepsilon(r)$, $p_{\varphi}(r)$ and $p_{r}(r)$ correspond to the energy density and tangential pressures of the shell in an orthonormal frame.
From the expression \eqref{eq:S}, the energy conditions that we focus on in this paper can be written as in Table.~\ref{table:EC}.

\begin{table}[h]
 \caption{Energy conditions}
 \label{table:EC}
 \centering
  \begin{tabular}{c|c}
   \hline \hline
   Null energy condition &  $\varepsilon+p_{\varphi}\geq0$, \hspace{2pt} $\varepsilon+p_{r}\geq0$  \\
   \hline 
   Weak energy condition & $\varepsilon+p_{\varphi}\geq0$, \hspace{2pt} $\varepsilon+p_{r}\geq0$, \hspace{2pt} $\varepsilon\geq0$  \\
   \hline 
   Dominant energy condition & $\varepsilon\geq|p_{\varphi}|$, \hspace{2pt} $\varepsilon\geq|p_{r}|$, \hspace{2pt}
   $\varepsilon\geq0$  \\
   \hline \hline
  \end{tabular}
\end{table}

By using Eqs.~\eqref{eq:Ktau}-\eqref{eq:Ktr}, the second junction condition can be expressed as follows:\\
\begin{widetext}
Eq.~\eqref{eq:2nd1}:
\begin{align}
\varepsilon(r) &= \frac{1}{8 \pi G} \left[
\nabla_{n} ( 2 U - \gamma )
+ \frac{\epsilon \sqrt{f(r)}}{ a(r)} \left( - \frac{Z'(r)}{R(r)} + \frac{Z'(r) R''(r) - R'(r) Z''(r)}{R'(r)^2 + Z'(r)^2}\right)
\right]^{-}, \label{eq:2ndep} \\
 p_{\varphi}(r) &=
\frac{1}{8 \pi G} \left[
\left( \nabla_{n} \gamma - \frac{\epsilon \sqrt{f(r)}}{a(r)} \frac{Z'(r) R''(r) - R'(r) Z''(r)}{R'(r)^2 + Z'(r)^2} \right)
\right]^{-}, \label{eq:2ndpphi} \\
p_{r}(r) &= \frac{1}{8 \pi G } \left[ \frac{\epsilon \sqrt{f(r)}}{a(r)} \frac{Z'(r)}{R(r)} \right]^{-}, \label{eq:2ndpr}
\end{align}    
Eq.~\eqref{eq:2nd2} :
\begin{align}
\biggl[
& - \left(\varepsilon(r)+ p_{\varphi}(r) + p_{r}(r) \right) \nabla_n U + p_{r}(r) \nabla_{n} \gamma \notag\\
&\qquad + \epsilon\frac{\sqrt{f(r)}}{a(r)} \left( p_{\varphi}(r) \frac{Z'(r)}{R(r)} - p_{r}(r) \frac{Z'(r) R''(r) - R'(r) Z''(r)}{R'(r)^2 + Z'(r)^2} \right)
\biggr]^{+} = 0,\label{eq:2ndKp}
\end{align}
Eq.~\eqref{eq:2nd3}:
\begin{align}
p_{r}'(r) + \frac{R'(r)}{R(r)} p_{r}(r)  + \left( \frac{f'(r)}{2 f(r)} - \frac{R'(r)}{R(r)}\right)p_{\varphi}(r) + \frac{f'(r)}{2 f(r)}\varepsilon(r)  = 0.\label{eq:2ndDS}
\end{align}
\end{widetext}
The last two equations \eqref{eq:2ndKp} and \eqref{eq:2ndDS} are automatically satisfied if the first three equations \eqref{eq:2ndep}, \eqref{eq:2ndpphi}, and \eqref{eq:2ndpr} are satisfied. Hence we will focus on these three equations.

\begin{figure*}[ht]
 \begin{minipage}{0.49\textwidth}
 \centering
 \includegraphics[width= \textwidth]{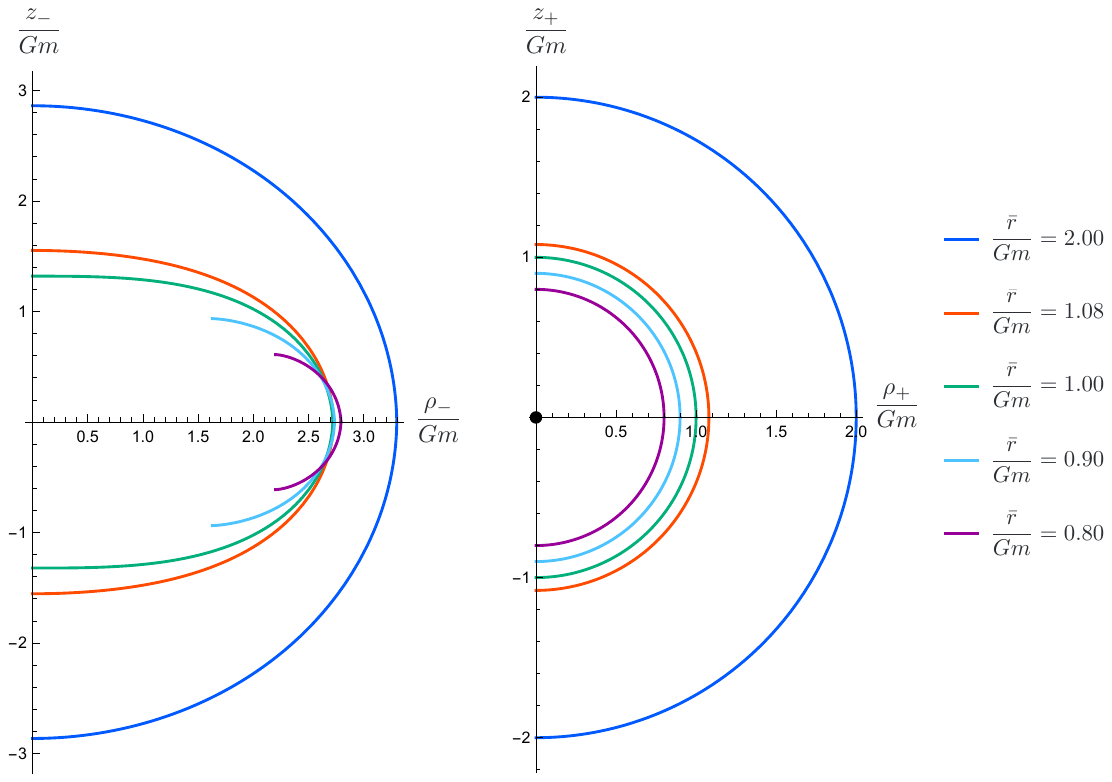}
 \caption{
Shell's position in the interior Weyl coordinates (left) and the exterior Weyl coordinates (right): The blue, red, green, cyan, and purple curves correspond to the parameter choice with $\bar{r}/G m=2.00, 1.08, 1.00, 0.90$ and $0.80$, respectively. The surface in the interior coordinates is closed only for $\bar{r}/Gm \geq 1$ (blue, red, and green).
The black dot in the right plot represents the curvature singularity.
}
    \label{fig:ccShell}
 \end{minipage}
\hfill
\begin{minipage}{0.49\textwidth}
\centering
\includegraphics[width= \hsize]{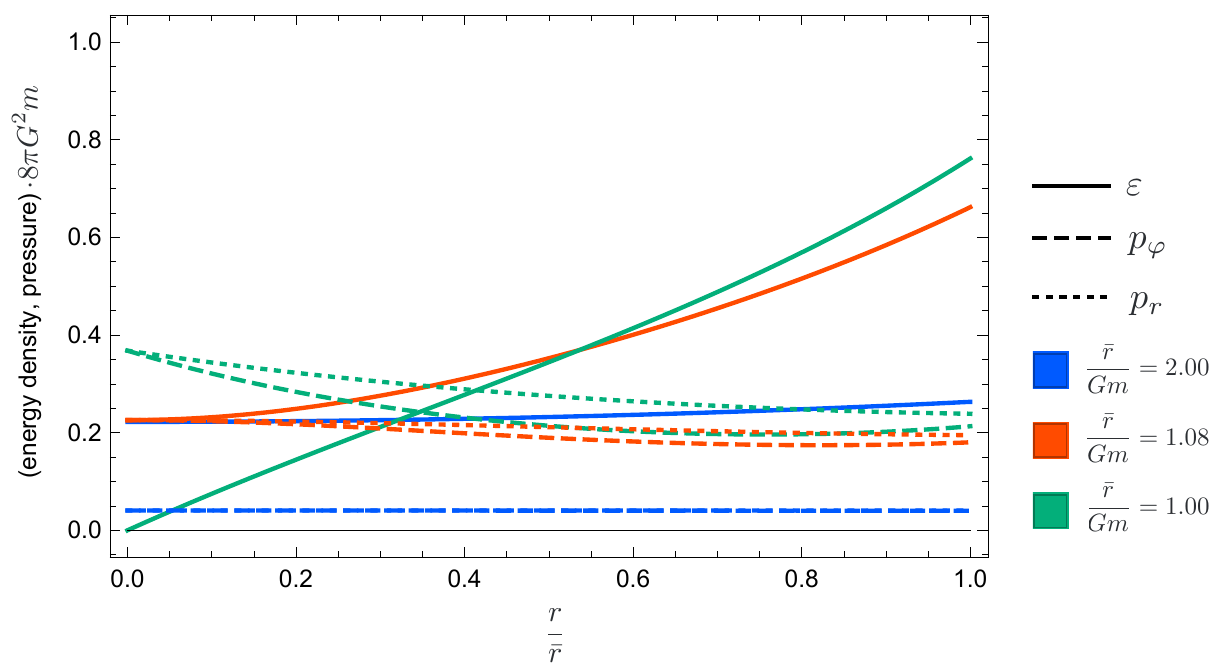}
\caption{Plot of $\varepsilon$ (solid), $p_{\varphi}$ (dashed), $p_{r}$ (dotted) as functions of $r/\bar{r}$.
The exterior geometry is fixed to be the Curzon--Chazy spacetime.
The blue, red, and green curves correspond to the cases with $\bar{r}/Gm=2.00, 1.08$ and $1.00$, respectively.
In the green plots, the pressures exceed the energy density for small $r$ and hence the dominant energy condition is violated, while, in the blue plots, the dominant energy condition holds for all $r$. The threshold is provided by the red plots. Since energy density and the pressures are positive for all cases, there is no violation of the weak and null energy conditions.}
\label{fig:epvsrMinCC}
\end{minipage}
\end{figure*}

\section{Explicit evaluation using exact solutions}
\label{exp}

In this section, we will explicitly evaluate the junction conditions by specifying the interior and exterior spacetimes. For our purpose of investigating non-singular spacetime, we assign the Minkowski spacetime as the interior. For the exterior we will focus on two simple cases within the Weyl class: a Curzon--Chazy spacetime and a Zipoy--Voorhees spacetime.

In the following, we will set $c_{+}=1$ and $R(r) = r$ without loss of generality.
In these parameter choices, the position of the shell in the exterior coordinates can be expressed simply as $\rho_{+} = r$ and $z_{+} = Z_{+}(r)$, while in the interior coordinates, it can be expressed as $\rho_{-}(r) = r/c_{-}$ and $z_{-}(r) = Z_{-}(r)/c_{-}$. The constants $\epsilon_{\pm}$ are set as $\epsilon_{\pm} = Z_{\pm}'(r)/|Z_{\pm}'(r)|$ as we identify the direction of $\boldsymbol{n}_{\pm}$ so that $\rho_{\pm}$ is increasing along $\boldsymbol{n}_{\pm}$: $n_{\pm}^{\mu} \nabla_{\mu} \rho_{\pm} = \theta_{n}{}^{\rho} > 0$.

\subsection{Curzon--Chazy spacetime}

Let us assume that the geometry of the two vacuum regions ${\cal M}^{+}$ and ${\cal M}^{-}$ are given by an outside of the Curzon--Chazy spacetime and an inside of the Minkowski spacetime, respectively.
As we have discussed in Sec.~\ref{SASS}, the Minkowski spacetime can be obtained by fixing the functions in Eq.~\eqref{eq:dsW} as
\begin{align}
&U_{-}=0, \quad \gamma_{-}=0,
\end{align}
and the Curzon--Chazy spacetime is obtained by 
\begin{align}
&U_{+}=-\frac{G m}{\sqrt{\rho^2_{+}+z_{+}^2}}, \quad \gamma_{+}=-\frac{G^2 m^2\rho_{+}^2}{2(\rho^2_{+}+z_{+}^2)^2},
\end{align}
where $m$ is the ADM mass.

In this system, the first junction conditions~\eqref{eq:1stf} and \eqref{eq:1sta} read
\begin{align}
& 
\mathrm{e}^{- \frac{2G m}{\sqrt{r^2+Z_{+}(r)^2}}} = c_{-}^2, \label{eq:CC1stf} \\
&\mathrm{e}^{-\frac{G^2 m^2 r^2}{2(r^2+Z_{+}(r)^2)^2}} \sqrt{1 + Z'_{+}(r)^2} = \sqrt{1 + Z'_{-}(r)^2}. \label{eq:CC1sta}
\end{align}
Here, we have used Eq.~\eqref{eq:1stR} with $R(r) = r$.
From Eq.~\eqref{eq:CC1stf}, we obtain
\begin{align}
&\sqrt{r^2+Z_{+}(r)^2} = \bar{r},
\label{eq:Zrp}
\end{align}
where we define a constant $\bar{r}$ by
\begin{align}
\bar{r} := - \frac{G m}{\log c_{-}}.\label{eq:defrbar}
\end{align}
Eq.~\eqref{eq:Zrp} is consistent only when $\bar{r} > 0$, which means that the parameter $c_{-}$ must satisfy $0 < c_{-} < 1$. 
Since the coordinate $r$ coincides with the radius of the Weyl coordinate $\rho_{+}$, Eq.~\eqref{eq:Zrp} indicates that, in terms of the Weyl coordinates for the exterior spacetime, the shape of the shell must be a sphere with the radius $\bar{r}$, as plotted in the right side of Fig.~\ref{fig:ccShell}.

By eliminating $Z_{+}(r)$ in Eq.~\eqref{eq:CC1sta}, we obtain
\begin{align}
Z'_{-}(r)^2 = \frac{\mathrm{e}^{ - \qty(\frac{Gm}{\bar{r}})^2\frac{r^2}{\bar{r}^2}}}{1 - \frac{r^2}{\bar{r}^2}} - 1.
\label{eq:Zmdash}
\end{align}
For $\bar{r} \geq G m $, the right-hand side is always positive for any $r \in (0,\bar{r})$. On the other hand, if $\bar{r} < G m$, the right-hand side is positive only for a finite range of $r \in (r_{*}, \bar{r})$, where $r_{*}$ is defined as the zero point of the right-hand side that is smaller than $\bar{r}$ and can be written as
\begin{align}
r_{*}=\bar{r}\sqrt{1+\qty(\frac{\bar{r}}{G m})^2 W\qty(-\qty(\frac{G m}{\bar{r}})^2e^{-\qty(\bar{r}/Gm)^2})},
\label{eq:rstar}
\end{align}
where $W(x)$ is Lambert's $W$ function. Hence we obtain a closed shell only when $\bar{r} \geq G m$.

By solving Eq.~\eqref{eq:Zmdash} numerically, we can express $Z_{+}(r)$ as a function of $r$. Hence, through the relation $\rho_{-} = r/c_{-}$ and $z_{-} = Z_{-}(r)/c_{-}$, as well as the relation between $\bar{r}$ and $c_{-}$ given by Eq.~\eqref{eq:defrbar}, the position of the shell in the inner Weyl coordinates $\{\rho_{-}, z_{-} \}$ can be obtained as plotted in the left side of Fig.~\ref{fig:ccShell}. 
From the figure, one can see that the shell can be closed only for $\bar{r} \geq G m$, as we have discussed above. We will focus only on such a case below.

Next, let us check the energy density and the pressures that are required through the second junction conditions. 
Since we choose the interior spacetime $\mathcal{M}_{-}$ as the Minkowski spacetime, in the second junction conditions, $Z_{-}(r)$ only appears through its derivatives $Z_{-}'(r)$ and $Z_{-}''(r)$. Hence, by using Eq.~\eqref{eq:Zmdash}, we can express the energy density and the pressures analytically as follows:
\begin{widetext}
\begin{align}
     \varepsilon(r) &= \frac{\mathrm{e}^{- \frac{Gm}{\bar{r}}+\frac{1}{2}\qty(\frac{r}{Gm})^2}}{8 \pi G^2 m}\frac{Gm}{\bar{r}} \left( - 2 + 2\frac{Gm}{\bar{r}} - \qty(\frac{Gm r}{\bar{r}^2})^2 + \frac{\bar{r}}{r}\sqrt{\mathrm{e}^{-\qty(\frac{Gm r}{\bar{r}^2})^2} - \qty(1 - \qty(\frac{r}{\bar{r}})^2)}  + \frac{r}{\bar{r}}\frac{ 1 - \frac{1 - (r/\bar{r})^2}{(\bar{r}/Gm)^2}}{\sqrt{\mathrm{e}^{-\qty(\frac{Gm r}{\bar{r}^2})^2} - \qty(1 - \qty(\frac{r}{\bar{r}})^2)}} \right), 
    \end{align} 
     \begin{align}
     p_{\varphi}(r) &= \frac{\mathrm{e}^{- \frac{Gm}{\bar{r}}+\frac{1}{2}\qty(\frac{r}{Gm})^2}}{8 \pi G^2 m}\frac{Gm}{\bar{r}} \left( 1 + \qty(\frac{Gm r}{\bar{r}^2})^2 - \frac{r}{\bar{r}}\frac{ 1 - \frac{1 - (r/\bar{r})^2}{(\bar{r}/Gm)^2}}{\sqrt{\mathrm{e}^{-\qty(\frac{Gm r}{\bar{r}^2})^2} - \qty(1 - \qty(\frac{r}{\bar{r}})^2)}} \right), \\
    p_r(r) &= \frac{\mathrm{e}^{- \frac{Gm}{\bar{r}}+\frac{1}{2}\qty(\frac{r}{Gm})^2}}{8 \pi G^2 m}\frac{Gm}{\bar{r}}
    \left(
    1 - \frac{\bar{r}}{r}\sqrt{\mathrm{e}^{-\qty(\frac{Gm r}{\bar{r}^2})^2} - \qty(1 - \qty(\frac{r}{\bar{r}})^2)}
    \right).
\end{align}    
\end{widetext}

The plots of them are shown in Fig.~\ref{fig:epvsrMinCC}.
The numerical plots show that $\varepsilon$, $p_{\varphi}$ and $p_{r}$ are always positive, and hence the energy-momentum tensor of the shell satisfies the weak and null energy condition (See Table.~\ref{table:EC}).
Note that the positivity of $\varepsilon$, $p_{\varphi}$ and $p_{r}$ can be shown analytically by using the inequalities
\begin{align}
&\sqrt{\mathrm{e}^{-\qty(\frac{Gm r}{\bar{r}^2})^2} - \qty( 1 - \qty(\frac{r}{\bar{r}})^2)}\geq \sqrt{1-\qty(\frac{Gm}{\bar{r}})^2}\frac{r}{\bar{r}}, \label{eq:ineq1} \\
&\frac{1}{\sqrt{\mathrm{e}^{-\qty(\frac{Gm r}{\bar{r}^2})^2} - \qty(1 - \qty(\frac{r}{\bar{r}})^2)}}\geq\frac{\bar{r}}{r}\sqrt{\frac{1+\qty(\frac{Gm r}{\bar{r}^2})^2}{1-\frac{1 - \qty(\frac{r}{\bar{r}})^2}{(\bar{r}/Gm)^2}}},  \label{eq:ineq2} 
\end{align}
which can be derived from the basic inequalities
\begin{align}
 1-\qty(\frac{Gm r}{\bar{r}^2})^2\leq e^{-\qty(\frac{Gm r}{\bar{r}^2})^2}\leq\frac{1}{1+\qty(\frac{Gm r}{\bar{r}^2})^2}.
\end{align} 

Fig.~\ref{fig:epvsrMinCC} also shows that the pressures exceed the energy density for a sufficiently small shell like the green plot. Hence the dominant energy condition is violated for a small shell. The threshold can be evaluated analytically as follows. 
First, for a fixed value of $ \bar{r} \geq Gm$, one can observe from Fig.~\ref{fig:epvsrMinCC} that the $\varepsilon - p_{\varphi}$ and $\varepsilon - p_{r}$ are increasing functions of $r$. We note that this property can also be shown analytically by using relations like Eqs.~\eqref{eq:ineq1} and \eqref{eq:ineq2}.
Then the functions $\varepsilon - p_{\varphi}$ and $\varepsilon - p_{r}$ take a minimum at $r = 0$, which can be evaluated as
\begin{align}
&\varepsilon(0) - p_{\varphi}(0) = \varepsilon(0) - p_{r}(0) \notag \\
&\quad  = \frac{1}{8 \pi G^2 m} \mathrm{e}^{-\frac{G m}{\bar{r}}} \left( \frac{2 - 3 \frac{\bar{r}}{G m} }{\qty(\frac{\bar{r}}{G m})^2} + 3 \sqrt{\qty(\frac{\bar{r}}{G m})^2 - 1} \right).
\end{align}
The right hand side is positive if and only if $\bar{r} \geq \bar{r}_{\textmd{th}} := \frac{13}{12}Gm\sim 1.08 Gm$.
Hence if $\bar{r} \geq \bar{r}_{\textmd{th}}$, the functions $\varepsilon(r) - p_{\varphi}(r)$ and $\varepsilon(r) - p_{r}(r)$ are positive for any $r \in [0,\bar{r}]$. In this case, the dominant energy condition holds.
On the other hand, if $\bar{r} < \bar{r}_{\textmd{th}}$, the functions at $r = 0$, $\varepsilon(0) - p_{\varphi}(0) = \varepsilon(0) - p_{r}(0) < 0$, become negative and the dominant energy condition is violated at least at $r = 0$.

\subsection{Zipoy--Voorhees spacetime}

Let us focus on the case where the exterior spacetime is the Zipoy--Voorhees spacetime, where the functions $U_{+}$ and $\gamma_{+}$ are given by \eqref{eq:UZV} and \eqref{eq:gammaZV} respectively.
We shall assume that the geometry inside of the shell is given by the Minkowski spacetime, as in the previous section.

Under this assumption, the first junction condition~\eqref{eq:1stf} reads
\begin{align}
    \frac{\sqrt{r^2 + (Z_{+}(r) - \ell)^2} + Z_{+}(r) - \ell}{\sqrt{r^2 + (Z_{+}(r) + \ell)^2} + Z_{+}(r) + \ell } =  c_{-}^{\frac{2 \ell}{G m}} =: c. \label{eq:defcZV}
\end{align}
One can see that there is no consistent solution for $c = 1$. In addition, by using the symmetry on $\ell \rightarrow - \ell$, one can always assume $0 < c < 1$ without loss of generality.

By squaring Eq.~\eqref{eq:defcZV}, we obtain the following expression
\begin{align}
&\sqrt{r^2 + (Z_{+}(r) \mp \ell)^2 } \nonumber \\
&=\pm \ell - z - \frac{r^2}{2} \frac{1 - c^{\pm2}}{(1 - c^{\pm 1}) z \mp (1 + c^{\pm 1}) \ell} \geq 0, \label{eq:consistency1}
\end{align}    
which provides consistency conditions.
Further, by squaring either of them, we obtain
\begin{align}
    \frac{Z_{+}^2}{\bar{z}^2} +\frac{r^2}{\bar{r}^2} = 1, \label{eq:ZVz+}
\end{align}
where
\begin{align}
\bar{z} &:= \frac{1 + c}{1 - c} \ell,
\qquad
 \bar{r} := \frac{2 \sqrt{c}}{1 - c} \ell.\label{eq:defzbar}
\end{align}
Thus we obtain that the shape of the shell must be an ellipse with semi-axes $\bar{z}$ and $\bar{r}$ in the exterior Weyl coordinates $\{\rho_{+}, z_{+}\}$, as shown in the right sides of Figs.~\ref{fig:zpShelll15}, \ref{fig:zpShelll10} and \ref{fig:zpShelll08} below.

Note that by solving~\eqref{eq:defcZV} with Eq.~\eqref{eq:defzbar}, $c_{-}$ can be expressed by $\ell$ and $\bar{r}$
\begin{align}
  c_{-} &= \qty(1 + 2\qty(\frac{ \ell}{\bar{r}})^2 \left( 1 - \sqrt{1 + \qty(\frac{\bar{r}}{\ell})^2}\right) )^{\frac{Gm}{2 \ell}}.
\end{align}
Similarly, the parameters $\bar{z}$ can be expressed as 
\begin{align}
 \bar{z} &= \sqrt{\ell^2 + \bar{r}^2}.
\end{align}
Hence $\bar{z}$ is larger than $\bar{r}$ and $\ell$, and the parameter $\ell$ measures a deviation from the round circle. Note that the relation $\bar{z} > \ell$ indicates that the naked singularity located at $\rho_{+} = 0, z_{+} \in [-\ell, \ell]$ is hidden by the shell for any choice of $\bar{r}$.

The consistency inequalitiy \eqref{eq:consistency1} reduces to $\bar{z}^2 \geq \ell |Z_{+}|$ which is automatically satisfied because $\bar{z} \geq |Z_{+}|$ and $\bar{z} \geq \ell$.

The other first junction condition \eqref{eq:1sta} can be expressed as 
\begin{align}
    Z'_{-}(r)^2 = 
    \frac{1}{1 - \frac{r^2}{\bar{r}^2}} \left(
    1 + \frac{\ell^2}{\bar{r}^2} \frac{r^2}{\bar{r}^2}
    \right)^{1 -\frac{G^2 m^2}{\ell^2}} - 1.
\end{align}
To obtain the solution, the right-hand side must be non-negative for any $r \in (0, \bar{r})$, which is achieved only when the parameters satisfy
\begin{align}
    G m \leq \sqrt{\bar{r}^2 + \ell^2} = \bar{z}. \label{eq:IneqGm}
\end{align}
One can plot the shape of the shell in the interior coordinates by integrating this equation numerically, as plotted in the left sides in Figs.~\ref{fig:zpShelll15}, \ref{fig:zpShelll10} and \ref{fig:zpShelll08}.

As similar to the previous subsection, the second junction conditions provide the analytic expression for $\varepsilon, p_{\varphi}$ and $p_{r}$ as a function of $r$. The results are given as follows:
\begin{widetext}
\begin{align}
\varepsilon(r) &= -p_{\varphi}(r) -p_{r}(r) + \frac{2}{8 \pi G^2 m} \frac{Gm}{\bar{r}} \left(1 + \qty(\frac{\ell}{\bar{r}})^2 \qty(\frac{r}{\bar{r}})^2\right)^{-\frac{1}{2} + \frac{G^2 m^2}{2 \ell^2}} \notag \\
&\qquad 
\times \qty(1+2\qty(\frac{\ell}{\bar{r}})^2-2\frac{\ell}{\bar{r}}\sqrt{1+\qty(\frac{\ell}{\bar{r}})^2})^{\frac{Gm}{2\ell}}\qty(\sqrt{1+\qty(\frac{\ell}{\bar{r}})^2}r\partial_{\rho_{+}}U_{+}-\sqrt{1 - \qty(\frac{r}{\bar{r}})^2 }\bar{r}\partial_{Z_{+}}U_{+}), \label{eq:epsilonZV} \\
p_{\varphi}(r) &=
\frac{1}{8 \pi G^2 m} \frac{Gm}{\bar{r}} 
\left( 1 + 2 \qty(\frac{\ell}{\bar{r}})^2 -2 \frac{\ell}{\bar{r}}\sqrt{1 + \qty(\frac{\ell}{\bar{r}})^2} \right)^{\frac{Gm}{2 \ell}}
\left(1 + \qty(\frac{\ell}{\bar{r}})^2 \qty(\frac{r}{\bar{r}})^2 \right)^{-\frac{3}{2} + \frac{G^2 m^2}{2 \ell^2}} \notag\\
&\qquad 
\times
\qty( \sqrt{1 + \qty(\frac{\ell}{\bar{r}})^2} \left(1 + \qty(\frac{Gm r}{\bar{r}^2})^2 \right) - \frac{\frac{r}{\bar{r}} \qty(1 + \qty(\frac{\ell}{\bar{r}})^2 - \qty(\frac{G m}{\bar{r}})^2 \left( 1 - \qty(\frac{r}{\bar{r}})^2 \right) )}{\sqrt{\left(1 + \qty(\frac{\ell}{\bar{r}})^2 \qty(\frac{r}{\bar{r}})^2 \right)^{1 - \frac{G^2 m^2}{\ell^2}} - \left(1 - \qty(\frac{r}{\bar{r}})^2 \right)}}), \label{eq:pphiZV}
\end{align}
and
\begin{align}    
p_{r}(r) &= \frac{1}{8 \pi G^2 m} \frac{G m}{\bar{r}}
\left(1 + 2 \qty(\frac{\ell}{\bar{r}})^2
-2 \frac{\ell}{\bar{r}}\sqrt{1 + \qty(\frac{\ell}{\bar{r}})^2}
\right)^{\frac{G m}{2\ell}} \left(1 + \qty(\frac{\ell}{\bar{r}})^2 \qty(\frac{r}{\bar{r}})^2 \right)^{ - \frac{1}{2} + \frac{G^2 m^2}{2\ell^2}} \notag\\
& \qquad \times
\left(
\sqrt{1 + \qty(\frac{\ell}{\bar{r}})^2}
- \frac{\bar{r}}{r} \sqrt{\left(1 + \qty(\frac{\ell}{\bar{r}})^2 \qty(\frac{r}{\bar{r}})^2 \right)^{1 - \frac{G^2 m^2}{\ell^2}} - \left(1 - \qty(\frac{r}{\bar{r}})^2 \right)}
\right), \label{eq:prZV}
\end{align}    
\end{widetext}
The plots of $\varepsilon, p_{\varphi}$ and $p_{r}$ are shown in Figs.~\ref{fig:NECvsrMinVZ1}, \ref{fig:NECvsrMinVZ2} and \ref{fig:NECvsrMinVZ3} for various choice of $\ell$.

Below, we will discuss the cases $\ell > Gm, \ell = Gm$ and $\ell < G m$ separately.  
\subsubsection{$\ell > G m$}
\begin{figure*}[htbp]
\begin{minipage}{0.49\textwidth}
\includegraphics[width= \hsize]{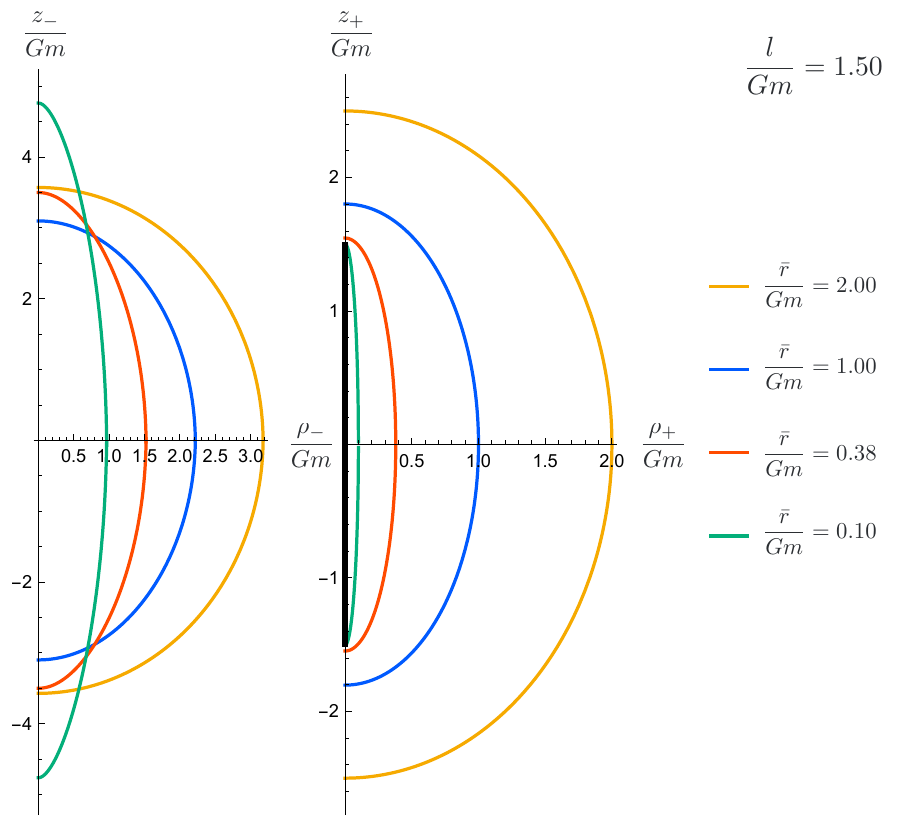}
    \caption{Shell's position in the interior Weyl coordinates (left) and the exterior Zipoy--Voorhees spacetime in the Weyl coordinates (right) for $\ell/Gm=1.50$: The orange, blue, red and green curves correspond to the parameter choice with $\bar{r}/G m=2.00, 1.00, 0.40$ and $0.10$, respectively. 
    The surface is closed for any value of $\bar{r}$.
The black vertical line in the right plot represents the curvature singularity.}
    \label{fig:zpShelll15}
\end{minipage}
\hfill
\begin{minipage}{0.49\textwidth}
\centering
\includegraphics[width= \hsize]{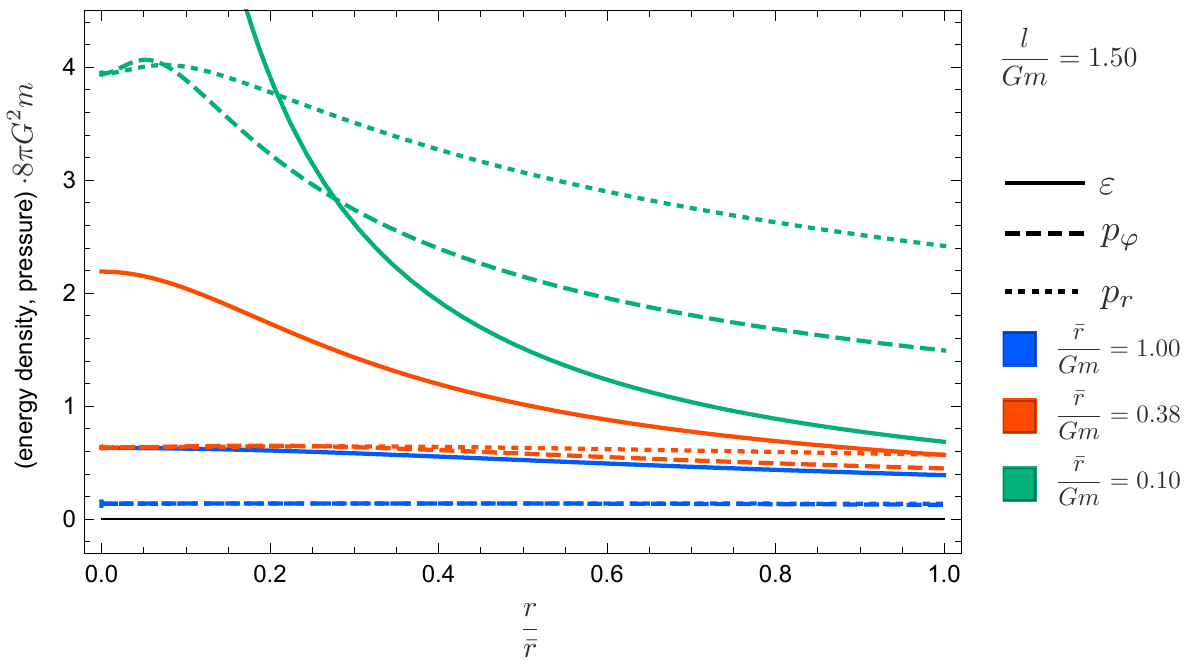}
\caption{Plot of $\varepsilon$ (solid), $p_{\varphi}$ (dashed), $p_{r}$ (dotted) as functions of $r/\bar{r}$.
The exterior geometry is fixed to be the Zipoy--Voorhees spacetime and we fix $\ell/Gm=1.50$.
The blue, red, and green curves correspond to the cases with $\bar{r}/Gm=2.00, 0.38$ and $0.10$, respectively.
In the green plots, the pressures exceed the energy density for small $r$ and hence the dominant energy condition is violated, while, in the blue plots, the dominant energy condition holds for all $r$. The threshold is provided by the red plots $(\bar{r}/Gm=0.38)$. Since energy density and the pressures are positive for all cases, there is no violation of the weak and null energy conditions.}
\label{fig:NECvsrMinVZ1} 
\end{minipage}
\end{figure*}

To begin with, let us focus on the case $\ell > Gm$. The numerical plots of the shape of the shells are given in Fig.~\ref{fig:zpShelll15} for $\ell = 1.5 Gm$. The plot of the energy density and pressures are shown in Fig.~\ref{fig:NECvsrMinVZ1}.

For $\ell > Gm$, the inequality \eqref{eq:IneqGm} is automatically satisfied for any $\bar{r}$. Hence there is no lower bound for the shell size and one can consider a shell arbitrarily close to the singularity. This property can be seen in the left side of Fig.~\ref{fig:zpShelll15}, where all shells are closed in the interior Weyl coordinates.

From Fig.~\ref{fig:NECvsrMinVZ1}, we find that $\varepsilon$, $p_{\varphi}$ and $p_{r}$ are always positive, which confirms that the weak and null energy conditions are satisfied.

Regarding the validity of the dominant energy condition, there is a lower bound on $\bar{r}$, $\bar{r}_{\textmd{th}}$. For example, from Fig.~\ref{fig:NECvsrMinVZ1}, one can see that $\bar{r}_{\textmd{th}} = 0.42$ for $\ell = 1.5 Gm$.
For the general choice of $\ell$, the threshold $\bar{r}_{\textmd{th}}$ can be obtained as follows.
First, one can numerically confirm that the functions $\varepsilon - p_{\varphi}$ and $\varepsilon - p_{r}$ are decreasing functions of $r$ for $\ell > G m$. Then the validity of the dominant energy conditions can be evaluated by checking the non-negativity of $\varepsilon - p_{\varphi}$ and $\varepsilon - p_{r}$ at $r = \bar{r}$. Then, the threshold $\bar{r}_{\textmd{th}}$ can be obtained as the larger root of $\varepsilon(\bar{r}) - p_{\varphi}(\bar{r})$ and $\varepsilon(\bar{r}) - p_{r}(\bar{r})$.
The numerical plot of $\bar{r}_{\textmd{th}}$ for $\ell > Gm$ case, as well as that for $\ell < G m$ which we will derive below, is shown in Fig.~\ref{fig:rth}.

We can also evaluate the proper length of the closed curve on the shell with $\bar{r}=\bar{r}_{\textmd{th}}$.
In Fig.~\ref{fig:Length}, the numerical plot of $L$, which is defined as the proper length devided by $2\pi$.
The red and blue curves denote the length of the $z$-constant curve and $\phi$-constant curve, respectively.
We observe that the length in the $z$-constant curve monotonically decreases with $\ell$. 
On the other hand, the length in the $\phi$-constant curve monotonically increases for $\ell> G m$.

\subsubsection{$\ell = G m$}
\begin{figure*}[htbp]
\begin{minipage}{0.49\textwidth}
\centering
\includegraphics[width= \textwidth]{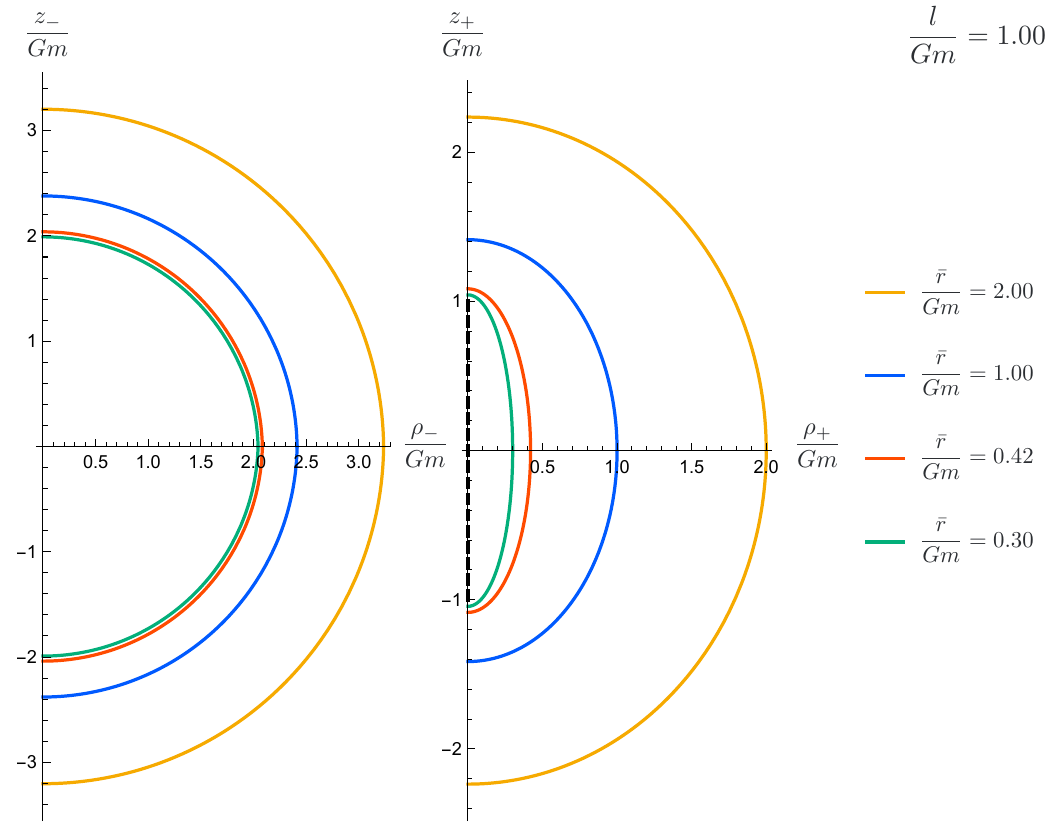}
    \caption{Shell's position in the interior Weyl coordinates (left) and the exterior Weyl coordinates (right) for $\ell/Gm=1.00$: The orange, blue, red and green curves correspond to the parameter choice with $\bar{r}/G m=2.00, 1.00, 0.42$ and $0.30$, respectively.
    In this case, the exterior geometry is given by the Schwarzschild spacetime and the shell is a spherical shape in the Weyl coordinates.
    The surface is closed for any value of $\bar{r}$.
The black vertical dashed line in the right plot represents the coordinate singularity.}
    \label{fig:zpShelll10}
\end{minipage}
\hfill
\begin{minipage}{0.49\textwidth}
\centering
\includegraphics[width= \textwidth]{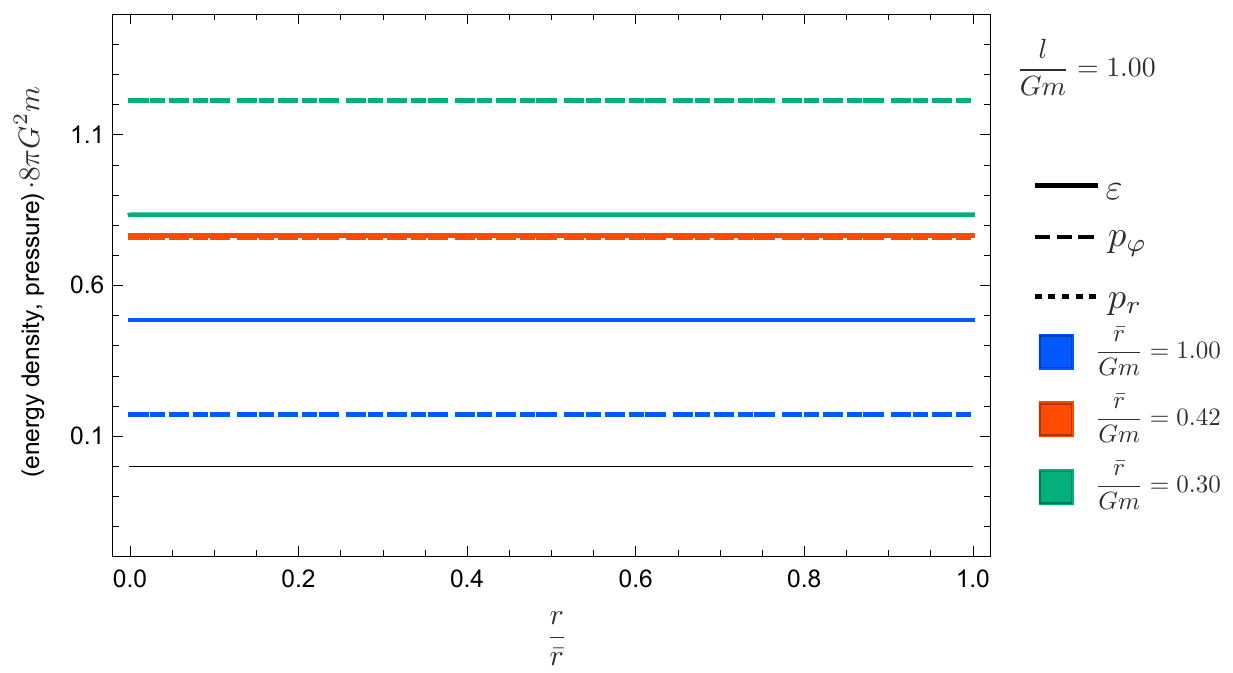}
\caption{Plot of $\varepsilon$ (solid), $p_{\varphi}$ (dashed), $p_{r}$ (dotted) as functions of $r/\bar{r}$ with $\ell/Gm=1.0$.
The blue, red, and green lines correspond to the cases with $\bar{r}/Gm=1.00, 0.42$ and $0.30$, respectively.
The value of the energy and pressures are independent of $r$.
Because of the spherical symmetry, the values of $p_{\varphi}$ and $p_{r}$ are degenerated.
In the green plots, the pressures exceed the energy density and hence the dominant energy condition is violated, while, in the blue plots, the dominant energy condition holds. The threshold is provided by the red plots $\qty(\bar{r}/Gm=0.42)$, where all of the lines are degenerated. Since energy density and the pressures are positive for all cases, there is no violation of the weak and null energy conditions.}
\label{fig:NECvsrMinVZ2}
\end{minipage}
\end{figure*}

For $\ell = Gm$, the numerical plots of the shape of the shells are given in Fig.~\ref{fig:zpShelll10} and the plot of the energy density and pressures are shown in Fig.~\ref{fig:NECvsrMinVZ2}.

Similar to the $\ell > Gm$ case, the inequality \eqref{eq:IneqGm} is automatically satisfied for any $\bar{r}$. Hence there is no lower bound for the shell size and one can consider a shell arbitrarily close to the event horizon. Since $\ell = Gm$ means that the exterior geometry is the spherically symmetric Schwarzschild spacetime, the shape of the shells in the interior coordinates are sphere as one can see from Fig.~\ref{fig:zpShelll10}.

Due to the spherical symmetry, the energy density and the pressures are constant on the shell and the two tangential pressures are equivalent, as seen from Fig.~\ref{fig:NECvsrMinVZ2}. From the expressions \eqref{eq:epsilonZV} - \eqref{eq:prZV}, one can find that the zero point of $\varepsilon - p_{\varphi} = \varepsilon - p_{r}$ is obtained for $\bar{r} = 5 G m/12  \sim 0.42 G m$, which also numerically confirmed from Fig.~\ref{fig:NECvsrMinVZ2}.

Since $c_{-} = 1/5$ for $\bar{r}/\ell = 5/12$, the threshold radius $\bar{r}_{\textmd{th}}$ corresponds to the sphere with the radius $25 G m/12 $ in the interior Weyl (and hence Minkowski) coordinates. This result is consistent with the analysis for the Schwarzschild spacetime in Ref.~\cite{Frauendiener:1990nao}.

\subsubsection{$\ell < G m$}
\begin{figure*}[htbp]
\begin{minipage}{0.49\textwidth}
\centering
\includegraphics[width = \textwidth]{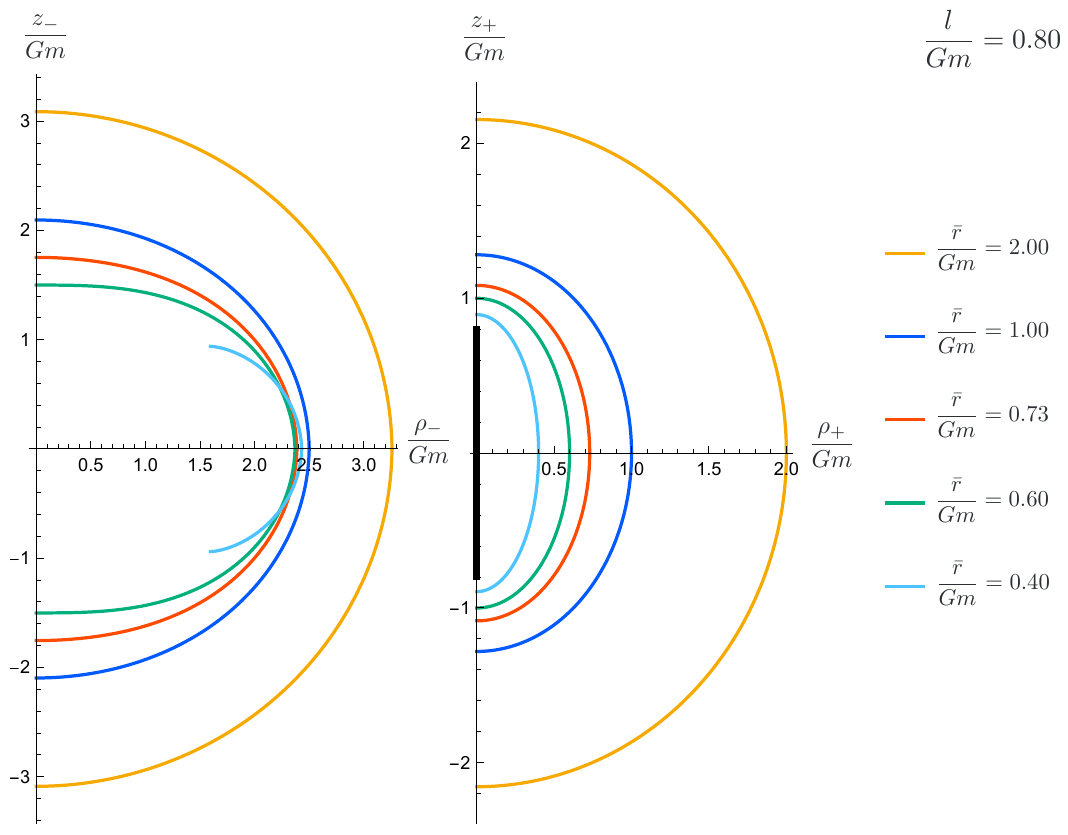}
    \caption{Shell's position in the interior Weyl coordinates (left) and the exterior Weyl coordinates (right) for $\ell/Gm=0.80$: The orange, blue, red, green and cyan curves correspond to the parameter choice with $\bar{r}/G m=2.00, 1.00, 0.73, 0.60$ and $0.40$, respectively.
    The surface in the interior coordinates is closed only for $\bar{r}/Gm \geq 0.60$ (orange, blue, red, and green).
The black vertical line in the right plot represents the curvature singularity.}
    \label{fig:zpShelll08} 
\end{minipage}
\hfill
\begin{minipage}{0.49\textwidth}
\centering
\includegraphics[width= \textwidth]{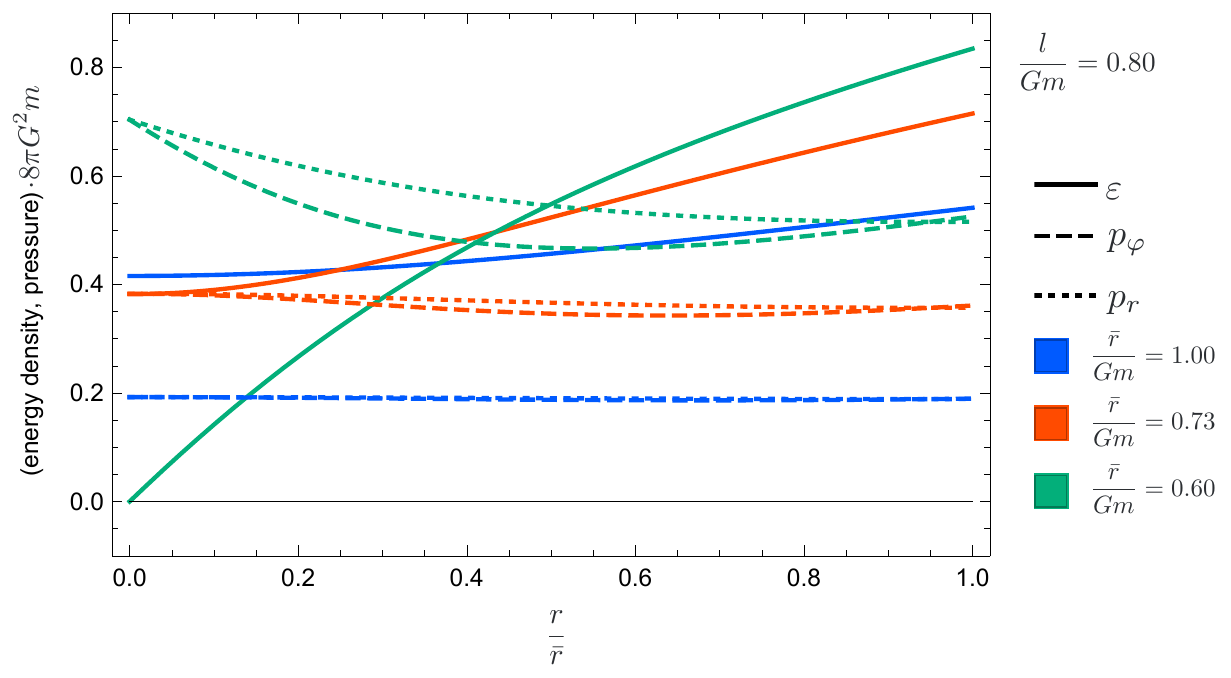}
\caption{Plot of $\varepsilon$ (solid), $p_{\varphi}$ (dashed), $p_{r}$ (dotted) as functions of $r/\bar{r}$ with $\ell/Gm=0.80$.
The blue, red, and green curves correspond to the cases with $\bar{r}/Gm=1.00, 0.73$ and $0.60$, respectively.
In the green plots, the pressures exceed the energy density for large $r$ and hence the dominant energy condition is violated, while, in the blue plots, the dominant energy condition holds for all $r$. The threshold is provided by the red plots $(\bar{r}/Gm=0.73)$. Since energy density and the pressures are positive for all cases, there is no violation of the weak and null energy conditions.}
\label{fig:NECvsrMinVZ3}
\end{minipage}
\end{figure*}

For $\ell < Gm$, the numerical plots of the shape of the shells are given in Fig.~\ref{fig:zpShelll08} and the plot of the energy density and pressures are shown in Fig.~\ref{fig:NECvsrMinVZ3}.

Contrary to the $\ell \geq Gm$ cases, the inequality \eqref{eq:IneqGm} is violated for small $\bar{r}$. Hence there is a lower bound for the shell size $\bar{r} = \bar{r}_{*}$ with
\begin{align}
 \bar{r}_{*} := \sqrt{G^2 m^2 - \ell^2}.\label{eq:rbarstarZV}
\end{align}
To form a closed shell, the parameter $\bar{r}$ must be greater than $\bar{r}_{*}$: $\bar{r} \geq r_{*}$.
For example, for $\ell = 0.8 Gm$, the minimum size of the shell is given as $\bar{r}_{*} = 0.6 Gm$. Since $c_{-} = (1/3)^{5/4}$ for the parameter choice $\bar{r}/\ell = 3/4$, this length $r = \bar{r}_{*}$ corresponds to $\rho_{-} = 0.6 \times 3^{5/4} Gm \sim 2.36 Gm$ in the interior coordinates. This can be confirmed also from the green plot on the left side of Fig.~\ref{fig:zpShelll08}, where the shell intersects with $\rho_{-}$ axes at $\rho_{-} = 2.36 Gm$.
    
Fig.~\ref{fig:NECvsrMinVZ3} shows the plots of the energy density and the pressures of the shell for $\ell = 0.8 Gm$. 
As similar to the other cases, we find that $\varepsilon$, $p_{\varphi}$ and $p_{r}$ are always positive, which confirms that the weak and null energy conditions are satisfied. 

\begin{figure*}[htbp]
\begin{minipage}{0.49\textwidth}
\includegraphics[width= \textwidth]{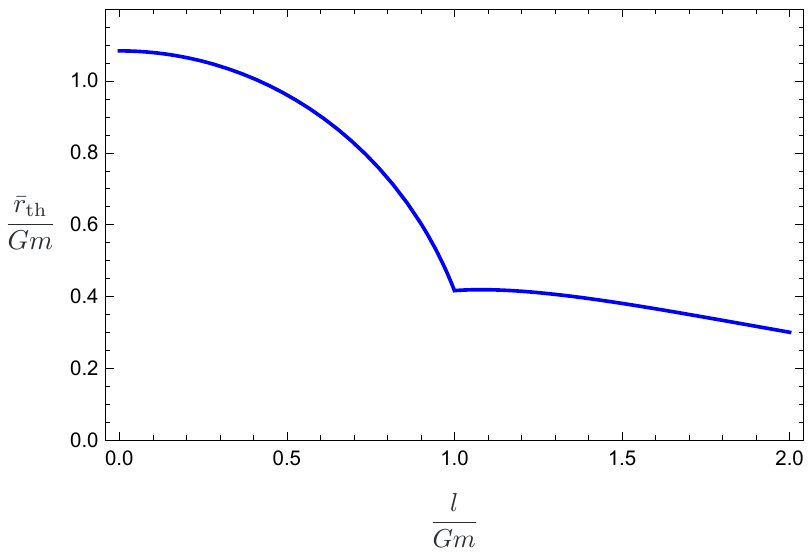}
\caption{Plot of the threshold value for $\bar{r}/Gm$ for the satisfaction of the dominant energy condition.
The threshold value $\bar{r}_{\textmd{th}}$ decreases monotonically with $\ell$.
The plot has a discontinuous derivative at $\ell=G m $.
This is because the violation of the dominant energy condition occurs in a small $r$ region for $\ell>G m $, while it occurs in a large $r$ region for $\ell<G m $.
For $\ell \leq G m $, the threshold can be obtained analytically as $\bar{r}_{\textmd{th}}=Gm\sqrt{\qty(\frac{13}{12})^2-\qty(\frac{\ell}{Gm})^2}$.
}
\label{fig:rth}
\end{minipage}
\hfill
\begin{minipage}{0.49\textwidth}
\centering
\includegraphics[width= \textwidth]{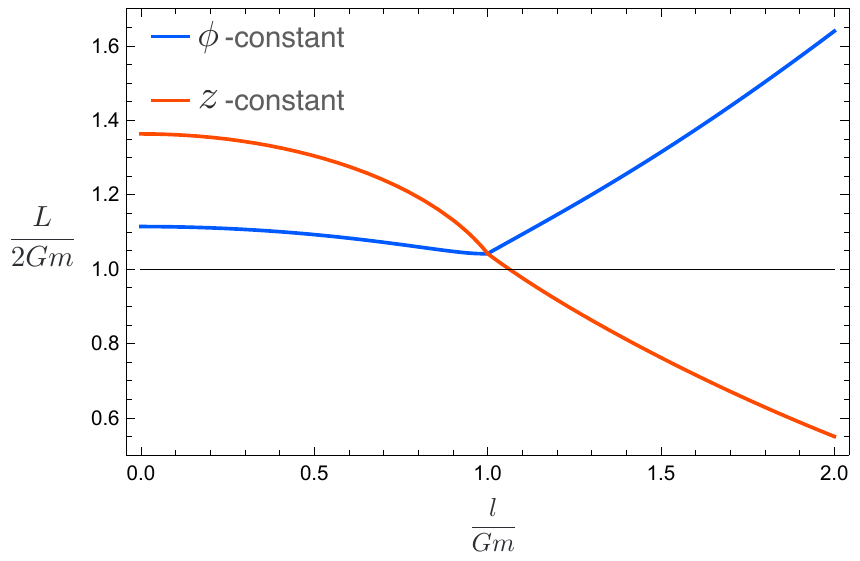}
\caption{$\ell$ dependence of the proper length along the shell with the minimal possible size $\bar{r}_{\textmd{th}}$.
The length $L$ is defined as the proper length of the closed curve divided by $2\pi$.
The blue and red curves correspond to the $\phi$-constant and the $z$-constant curves on the shell, respectively.
The red curve indicates the monotonic increase of the length with $\ell$, while the blue curve has a discontinuous derivative at $\ell=G m $.
The length of the $\phi$-constant curve satisfies $L>2 G m$ for any value of $\ell$.}
\label{fig:Length}
\end{minipage}
\end{figure*}

The dominant energy condition is violated only for small shells, such as the green plot ($\bar{r} = 0.6 Gm$) in Fig.~\ref{fig:NECvsrMinVZ3}.
The threshold for the shell size for the validity of the dominant energy condition can be obtained as follows. First one can numerically show that $\varepsilon - p_{\varphi}$ and $\varepsilon - p_{r}$ are increasing functions of $r$ and these minimum values are obtained at $r = 0$. Then one can see that $\varepsilon(0) - p_{\varphi}(0)$ and $\varepsilon(0) - p_{r}(0)$ vanish when $\bar{r} = \bar{r}_{\textmd{th}}$ that is given by 
\begin{align}
    \bar{r}_{\textmd{th}}=Gm\sqrt{\qty(\frac{13}{12})^2-\qty(\frac{\ell}{Gm})^2},
\end{align}
which is plotted in Fig.~\ref{fig:rth}. 
For example, $\bar{r}_{\textmd{th}}$ for $\ell = 0.8 Gm$ case can be evaluated as $\bar{r}_{\textmd{th}} = \sqrt{17\times113} G m/60 \sim 0.73 Gm$.

The proper length $L$ of the closed curves on the shell with $\bar{r}=\bar{r}_{\textmd{th}}$ is shown in Fig.~\ref{fig:Length}.
We observe that both lengths of the $z$-constant curve and $\phi$-constant curve decrease with $\ell$ for $\ell< G m$.

\section{Summary and Discussion}
\label{summary}

In this paper, we explored the construction of non-singular spacetimes with static and axial symmetry.
We focused on the class of solutions with asymptotic flatness, i.e., the Weyl class, and discussed how to glue spacetimes with a static shell.
We imposed the continuity of the induced metric (the first junction condition) to obtain the configuration of the shell and then evaluated the value of the components of the energy-momentum tensor from the difference in the extrinsic curvature (the second junction condition).
We conducted the explicit calculation by assuming that the geometry inside is given by the Minkowski spacetime and the outside is the Curzon--Chazy/Zipoy--Voorhees spacetimes.

The results of the Curzon--Chazy spacetimes can be summarized as follows:
We confirmed that a Curzon--Chazy spacetime with a positive ADM mass $m$ can be realized as an exterior of a spherical shell located at 
\begin{align}
    \rho_{+}^2 + z_{+}^2 = \bar{r}^2,
\end{align}
where the radius must be greater than $Gm$, $\bar{r} \geq Gm$, for the shell to be closed. 
Since the curvature singularity of the Curzon--Chazy spacetimes is located at the origin in the Weyl coordinates, the resultant spacetime is singularity-free.
If $\bar{r} \geq 13 G m/12$, the energy-momentum tensor of the shell is consistent with the weak, null, and dominant energy conditions. On the other hand, if $ Gm \leq \bar{r} < 13 G m/12$, the energy-momentum tensor of the shell respects the weak and null energy condition, while violating the dominant energy condition.

Also, we confirmed that a spacetime within the Zipoy--Voorhees class, with a parameter $\ell > 0$ and ADM mass $m > 0$, can be realized as an exterior of a shell located at
\begin{align}
 \frac{z_{+}^2}{\bar{r}^2 + \ell^2} + \frac{\rho_{+}^2}{\bar{r}^2} = 1,
\end{align}
in the Weyl coordinates for the exterior spacetime. The free parameter $\bar{r}$ characterizes the shell size. 
For $\ell < Gm$, the parameter $\bar{r}$ must satisfy $\bar{r} \geq \bar{r}_{*}$ where $\bar{r}_{*}$ is defined by Eq.~\eqref{eq:rbarstarZV}, while there is no minimum size for $\ell \geq Gm$. Since the curvature singularity in the Zipoy--Voorhees spacetime is hidden by the shell, the resultant spacetime is singularity-free. 
The weak and null energy conditions are satisfied for any shell size. However, to satisfy the dominant energy condition, the shell size must be greater than the threshold $\bar{r}_{\textmd{th}}$: $r \geq \bar{r}_{\textmd{th}}$. The threshold $\bar{r}_{\textmd{th}}$ is decreasing function on $\ell$, as shown in Fig.~\ref{fig:rth}.
The proper length $L$ of the closed curve on the shell with $r = \bar{r}_{\textmd{th}}$ cannot be smaller than the horizon radius of the Schwarzschild spacetime as shown in Fig.~\ref{fig:Length}.
That result can be attributed to the hoop conjecture~\cite{klauder1972magic,Misner:1973prb}.
According to the hoop conjecture, the apparent horizon forms if and only if the inequality $L\leq 2 G m$ is satisfied for any direction.
Since we have fixed the exterior geometry as the Zipoy--Vooheers spacetime which has a naked singularity, the horizon formation condition should be violated even if it is hidden by a thin shell.

In this paper, we have focused on patching static and axially symmetric spacetimes under certain assumptions.
Specifically, we have restricted ourselves to cases such that the geometry is pathed by a static shell and the interior geometry is given by the Minkowski solution.
Relaxing these assumptions may affect the condition for the shell to close and the satisfaction of the energy conditions.
For example, extending this approach to axially symmetric but stationary rotating spacetimes, such as those in the Tomimatsu--Sato class~\cite{Tomimatsu:1972zz,Tomimatsu:1973xvr}, is a potential direction for future research.
Modifying the geometry near the center, such as by introducing a stellar-like extended structure, would also be a valuable extension, which we leave for future work.

\section*{Acknowledgements}

This work was partially supported by JSPS KAKENHI Grant Numbers 24KJ1223 (DS) and JP20K14469 (DY). DY was also supported by Grant-Aid for Scientific Research from Ministry of Education, Science, Sports and Culture of Japan JSPS (No. JP21H05189).

\bibliography{ref}
\end{document}